\pdfoutput=1
\documentclass[%
 prl,
 amsmath,amssymb,
 preprint,
superscriptaddress,
author-year,
longbibliography
]{revtex4-2}

\usepackage{amsmath,bm,dcolumn,graphicx,xspace,tabularx, soul}
\usepackage{xcolor}
\usepackage{xr-hyper}

\makeatletter
\newcommand*{\addFileDependency}[1]{
  \typeout{(#1)}
  \@addtofilelist{#1}
  \IfFileExists{#1}{}{\typeout{No file #1.}}
}
\makeatother

\newcommand*{\myexternaldocument}[1]{%
    \externaldocument{#1}%
    \addFileDependency{#1.tex}%
    \addFileDependency{#1.aux}%
}

\myexternaldocument{revision_supplementary_3}

\setcitestyle{round}
\usepackage{totcount}
\newtotcounter{citnum} 
\def\oldbibitem{} \let\oldbibitem=\bibitem
\def\bibitem{\stepcounter{citnum}\oldbibitem}

\newcommand{\ii}{{i\mkern1mu}}
\makeatletter
\newcommand*\bigcdot{\mathpalette\bigcdot@{.5}}
\newcommand*\bigcdot@[2]{\mathbin{\vcenter{\hbox{\scalebox{#2}{$\m@th#1\bullet$}}}}}
\makeatother

\newcommand{\invFFT}{\hat{\mathcal{F}}^{-1}_{\mathbf{k}\to\mathbf{r}}}
\newcommand{\FFT}{\hat{\mathcal{F}}_{\mathbf{r}\to\mathbf{k}}}

\usepackage{etoolbox}
\patchcmd{\section}
  {\centering}
  {\raggedright}
  {}
  {}
\patchcmd{\subsection}
  {\centering}
  {\raggedright}
  {}
  {}
\usepackage{titlesec}
\titleformat{\section}
  {\normalfont\large\bfseries}{\thesection}{1em}{}[{\titlerule[0.8pt]}]

\emergencystretch=\maxdimen
\newcolumntype{L}[1]{>{\raggedright\let\newline\\\arraybackslash\hspace{0pt}}m{#1}}
\newcolumntype{C}[1]{>{\centering\let\newline\\\arraybackslash\hspace{0pt}}m{#1}}
\newcolumntype{R}[1]{>{\raggedleft\let\newline\\\arraybackslash\hspace{0pt}}m{#1}}

\usepackage{booktabs}

\usepackage{graphicx,dcolumn,amsmath,amssymb,color,mathrsfs,titlesec,hyperref,enumerate,bm,subfigure,float,comment}
\definecolor{linkColor}{rgb}{1,0,0}
\hypersetup{pdfborder={0 0 0},colorlinks=true,urlcolor=linkColor,citecolor=linkColor}
\usepackage{siunitx}

\renewcommand\thesection{\arabic{section}}

\begin{document}


\title{A robust synthetic data generation framework for machine learning in High-Resolution Transmission Electron Microscopy (HRTEM)}

\author{Luis Rangel DaCosta}
\email{luisrd@berkeley.edu}
\affiliation{Department of Materials Science and Engineering, University of California Berkeley, Berkeley, CA 94720}
\affiliation{National Center for Electron Microscopy, Molecular Foundry, Lawrence Berkeley National Laboratory, 1 Cyclotron Road, Berkeley, CA, USA}

\author{Katherine Sytwu}
\affiliation{National Center for Electron Microscopy, Molecular Foundry, Lawrence Berkeley National Laboratory, 1 Cyclotron Road, Berkeley, CA, USA}

\author{C.K. Groschner}
\affiliation{Department of Materials Science and Engineering, University of California Berkeley, Berkeley, CA 94720}
\affiliation{National Center for Electron Microscopy, Molecular Foundry, Lawrence Berkeley National Laboratory, 1 Cyclotron Road, Berkeley, CA, USA}

\author{M.C. Scott}
\email{mary.scott@berkeley.edu}
\affiliation{Department of Materials Science and Engineering, University of California Berkeley, Berkeley, CA 94720}
\affiliation{National Center for Electron Microscopy, Molecular Foundry, Lawrence Berkeley National Laboratory, 1 Cyclotron Road, Berkeley, CA, USA}

\date{\today}
\begin{abstract}

 Machine learning techniques are attractive options for developing highly-accurate automated analysis tools for nanomaterials characterization, including high-resolution transmission electron microscopy (HRTEM). However, successfully implementing such machine learning tools can be difficult due to the challenges in procuring sufficiently large, high-quality training datasets from experiments. In this work, we introduce Construction Zone, a Python package for rapidly generating complex nanoscale atomic structures, and develop an end-to-end workflow for creating large simulated databases for training neural networks. Construction Zone enables fast, systematic sampling of realistic nanomaterial structures, and can be used as a random structure generator for simulated databases, which is important for generating large, diverse synthetic datasets. Using HRTEM imaging as an example, we train a series of neural networks on various subsets of our simulated databases to segment nanoparticles and holistically study the data curation process to understand how various aspects of the curated simulated data---including simulation fidelity, the distribution of atomic structures, and the distribution of imaging conditions---affect model performance across several experimental benchmarks. Using our results, we are able to achieve state-of-the-art segmentation performance on experimental HRTEM images of nanoparticles from several experimental benchmarks and, further, we discuss robust strategies for consistently achieving high performance with machine learning in experimental settings using purely synthetic data.

\vspace{1.6cm}
\end{abstract}
\pacs{}
\keywords{Transmission Electron Microscopy, Machine Learning, Simulation, Open Source}
\maketitle

\section{Introduction}

Machine learning (ML) methods promise to accurately and automatically analyze large datasets at high-speeds, revolutionizing our materials characterization workflows. Many state-of-the-art ML tools rely on supervised learning techniques, where models utilize large amounts of data annotated with features of interest for training. The performance of supervised ML models, like neural networks, directly depends on the contents and generating distribution of the dataset used for model training, and, importantly, such models have been shown to extrapolate poorly beyond their training datasets \cite{xu_how_2021} and have limited out-of-distribution generalization behavior \cite{nagarajan_understanding_2021, krueger_out--distribution_2021}. Developing robust ML models for automated analysis of transmission electron microscopy (TEM), a versatile technique for structural and functional materials characterization at the atomic-scale, thus requires large image datasets which fully cover experimental imaging conditions and the variety of samples one has imaged. However, manually producing sufficiently large and diverse sets of well-annotated experimental data in order to train robust, generalizable ML models can be extremely labor intensive and creates the possibility for both human and experimental biases to negatively impact model performance during deployment. With limited experimental data, it is also difficult to investigate any failures or biases of a ML workflow arising from the choice of data used. 

The prohibitive cost of producing high-quality, well-annotated experimental data for supervised learning tasks makes data simulation an attractive alternative for developing effective machine learning models. Synthetic datasets produced through materials simulations offer several distinct advantages over their experimental counterparts. In particular, high-throughput simulation can create arbitrarily large datasets covering the full range of experimental conditions with ground-truth, physics-based data annotation, avoiding human bias and error in selecting and annotating relevant data and at a lower cost. Synthetic data generation also enables consistently reproducible end-to-end model development workflows and the ability to precisely isolate data stream effects---both positive and negative---on trained models. The final challenge then becomes choosing suitable and sufficient datasets for developing ML models to achieve a scientific task of interest, making it important to understand how the data we use influences the accuracy and quality of the scientific inferences we make and how such data curation decisions induce practical compromises between ML model performance and model development costs.

Recent advances in electron microscopy simulation methods have enabled large-scale, high-throughout TEM simulations for a wide range of experimental modalities \cite{rangel_dacosta_prismatic_2021, pryor2017streaming, ophus2017fast, abtem, lobato2015multem}. These simulation tools fully describe imaging with electron microscopy with detailed descriptions of the electron scattering process and have been used to train ML models to analyze crystalline scanning TEM (STEM) and STEM diffraction data \cite{munshi_disentangling_2022, zhang_atomic_2020}, high-resolution TEM (HRTEM) analysis of 2D material structures and nanoparticles \cite{madsen_deep_2018}, and image denoising models \cite{vincent_developing_2021}. However, these prior achievements either have limited types of atomic structures \cite{zhang_atomic_2020, madsen_deep_2018, vincent_developing_2021} or work only with crystalline diffraction data \cite{munshi_disentangling_2022}. In contrast, developing ML models general enough to analyze arbitrary real-space HRTEM data requires including all possible atomic structures that span the experimental range in the training datasets. Sampling many thousands of unique atomic structures and configurations with, for example, high-throughput molecular dynamics quickly becomes computationally expensive. At the expense of precise structural accuracy, we can instead cheaply generate thousands of atomic structures manually by specifying broad structural details of a nanoscale object and using idealized lattices and defect structures to place atoms. Generating such a set of structures for a bespoke problem or singular experiment is feasible, but, while several tools exist to design and analyze atomic structures in materials science \cite{ong_python_2013, larsen_atomic_2017, rahm_wulffpack_2020}, no such tools are designed to facilitate the precise description and generation of arbitrary distributions of complex, nanoscale atomic objects. Curating a sufficient dataset for training ML models to automatically and precisely analyze HRTEM characterization data, where one expects to encounter a vast diversity of highly numerous and variable atomic structures, becomes substantially more difficult in the general case without more flexible software tools.

\begin{figure}[hbtp]
    \centering
    \includegraphics[width=6in]{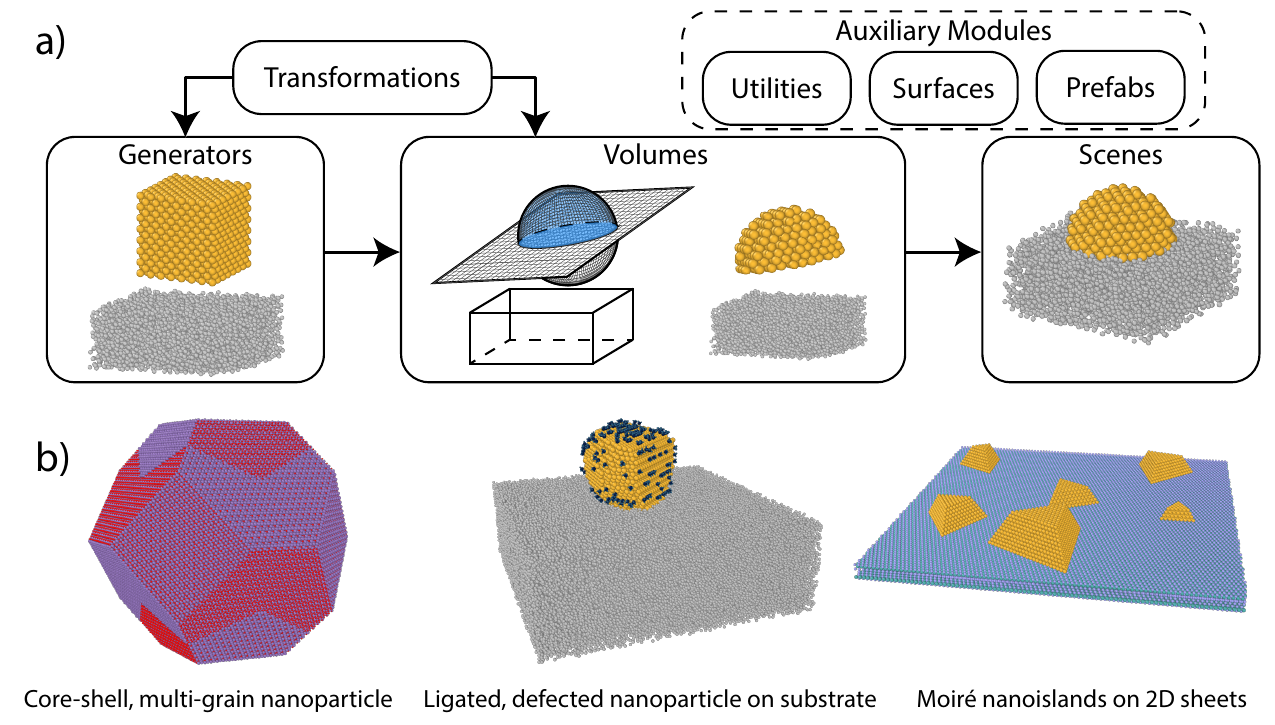}
    \caption{Diagram of the modular structure of Construction Zone (a). Atoms are supplied by Generator objects; subtractively removed into convex objects by Volume objects; and combined together into Scenes, in which multiple objects interact. A Transformations module provides both standard symmetry operations and more complex modifications to Generators and Volumes. The rest of the package includes miscellaneous utilities for interfacing with other software, premade structural archetypes, and tools for surface analysis and modification. (b) Example structures generated in Construction Zone, including a multi-grain core-shell oxide nanoparticle with strain-mediated grain alignment \cite{oh_design_2020} (left), a heavily-faceted gold nanoparticle on a carbon substrate decorated with molecule ligands (center), and a series of gold nanoislands on a bilayer of MoS$_2$ forming Moir\'e heterostructures \cite{reidy_direct_2021} (right)}
    \label{fig:cz_diagram}
\end{figure}

In this work, we develop an end-to-end workflow for training ML models to automatically analyze experimental HRTEM data of nanomaterials at the atomic-scale using only large, automatically generated, high-quality synthetic datasets. To achieve this, we develop Construction Zone (CZ) (Fig. \ref{fig:cz_diagram}), an open-source software package which enables algorithmic and high-throughput sampling of arbitrary atomic structures, which is then combined with HRTEM simulation to generate metadata rich databases with physics-based supervision labels. Given a general, all-purpose framework for producing high-quality synthetic data from a variety of prior distributions, we evaluate differing data curation strategies to reveal data-efficient training methods for achieving state-of-the-art neural network performance in analyzing experimental HRTEM data. Further, utilizing our data curation framework, which gives us complete control of the data generation process, we study statistically precise relationships between aspects of the training data---including simulation fidelity, structural composition, and diversity of imaging conditions---and ML model performance, aggregating the performance results of several hundreds of neural networks in order to provide statistical robustness. We provide analysis for image segmentation tasks, a benchmark characterization test which is crucial for identifying regions of interest in HRTEM data and guiding downstream automated analysis tasks, and present results on experimental HRTEM micrographs of clusters of Au and CdSe nanoparticles imaged at ultra-high magnification.

\section{Results}

\subsection{Atomic Structure and HRTEM Image Database Generation}

Models developed with ML methods for TEM analysis need to be adapted to the specific atomic structures one plans or expects to analyze in experiments; therefore, the training dataset should include examples of a wide number of likely atomic structures, so that any experimentally observed atomic structures lie close to the distribution of structures used for training. When curating a dataset to train robust HRTEM analysis models, one needs to be able to effectively and simultaneously capture both broad, high level aspects of atomic structures alongside the fine details that can be used to fully describe an individual structure. For example, an atomic structure may belong to a broader family of related structures, such as core-shell nanoparticles with similar sizes and chemistries. The same structure can also be described with unique and specific details, such as the placement of a defect plane or the orientation of its lattice with respect to the electron beam. To fully capture and utilize these complementary structural details, we have developed Construction Zone (CZ), an open-source Python package for building arbitrary atomic scenes at the nanoscale. CZ is designed to be robust to general use-cases, such that any complex nanostructure can be made, whilst also facilitating a flexible, programmatic workflow, such that large distributions of similar objects can be generated quickly and easily, and structure generating code is both easy to interpret and easy to reuse and repurpose. 

CZ relies on a simple module structure (c.f. Fig. \ref{fig:cz_diagram}a) that combines atomic placement (Generators), nano-object creation (Volumes), and nano-object interaction (Scenes). Structures can be further manipulated or generated with the Transformation class, which contains methods like standard symmetry operations, or by using convenience routines and analytical tools from the auxiliary modules, including functionality for atomic surface analysis and modification. The package derives some of its core functionality from other open-source materials science software packages, namely, PyMatgen \cite{ong_python_2013}, the Atomic Simulation Environment \cite{larsen_atomic_2017}, and WulffPack \cite{rahm_wulffpack_2020}, and interfaces seamlessly into common simulation workflows. By allowing users to specify atomic features like defects and zone-axis orientations with a high-level, materials-focused interface, CZ enables easier generation of both specific and random nanoscale atomic structures. We showcase some example complex atomic structures ranging from nanoparticles to 2D heterostructures, generated entirely with CZ, in Fig. \ref{fig:cz_diagram}b. In our study, CZ allows us to sample and quickly generate a large number of random, similar nanoparticle structures with complex defects and varying zone-axis orientations, mimicking the collection of nanoparticles that might be imaged in a typical HRTEM experiment, whilst also tracking such metadata about each structure, enabling us to draw fully specified training data distributions for machine learning model development.

Here, we use CZ as a random structure generator alongside high-throughput TEM simulation to create large, synthetic datasets which we use to train neural networks to analyze experimental HRTEM data, utilizing supervised learning techniques. We demonstrate our framework on the benchmark task of image segmentation for micrographs of Au and CdSe nanoparticles on amorphous carbon substrates \cite{groschner_machine_2021}. For each image, the goal is to classify each pixel as either part of a nanoparticle or substrate. Due to the subtle, complex interplay of contrast effects under HRTEM imaging at atomic resolution, classical image analysis techniques like Fourier filtering fail to segment nanoparticles accurately, and are outperformed by neural networks trained on manually-labeled experimental micrographs \cite{groschner_machine_2021}. Unsupervised techniques, such as using K-means clustering for pixel classification based on intensity, have only been successfully deployed for TEM images of nanoparticles taken at lower magnification \cite{wang_autodetect-mnp_2021}, where amplitude contrast dominates the signal, atomic lattice texture is not present, and data tend to be less noisy, thus further motivating high-accuracy supervised learning methods as a more successful route for image segmentation at atomic resolution.

\begin{figure}[hbtp]
    \centering
    \includegraphics[width=6in]{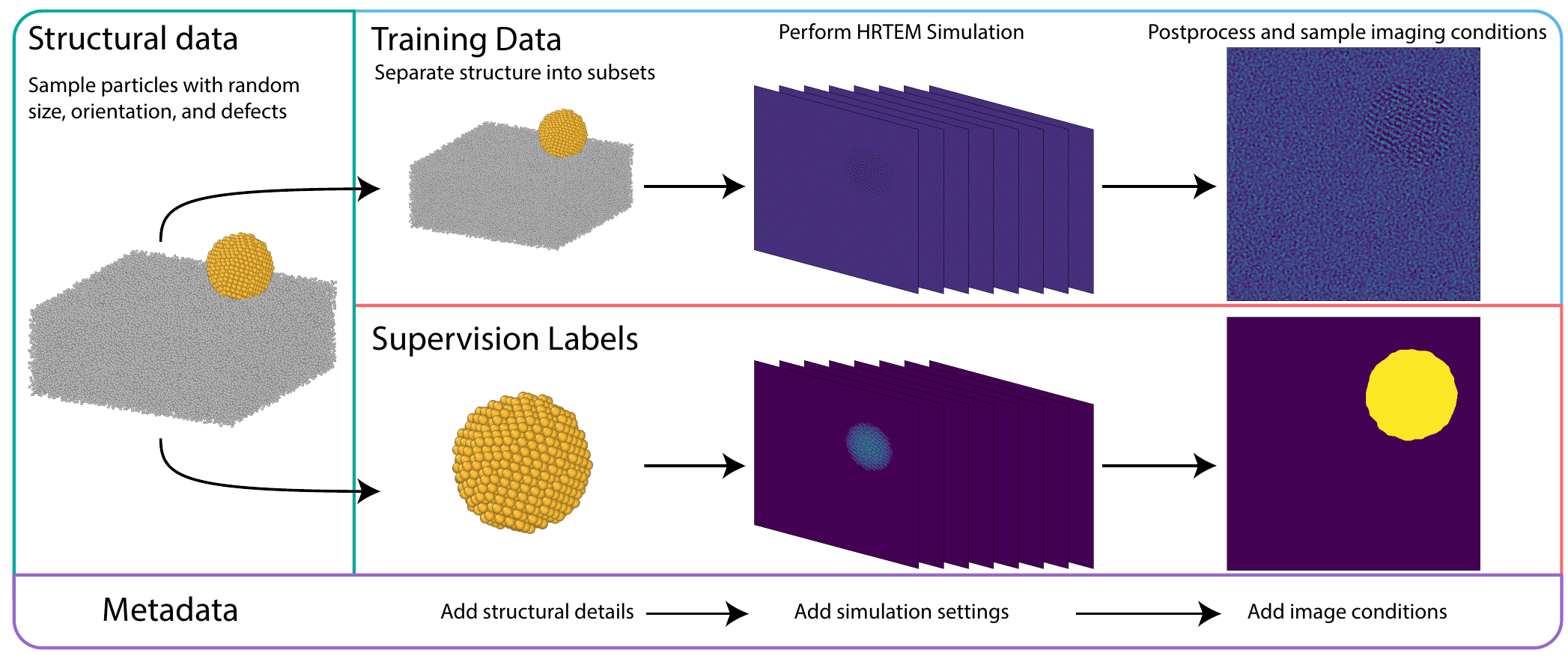}
    \caption{Data processing pipeline from generated structures to training-ready data. For each structure, we separate the structure into subsets for the training data and supervision label, simulate the HRTEM image formation for each structure with Prismatic \cite{rangel_dacosta_prismatic_2021}, and use post-processing to sample imaging conditions, noise, and generate the segmentation mask. Metadata are accrued at each step, and the data are stored into a series of staged databases.}
    \label{fig:data_pipeline}
\end{figure}

To generate our synthetic dataset for training neural networks, we built a data generation pipeline, diagrammed in Fig. \ref{fig:cz_diagram} and detailed more thoroughly in Fig. \ref{sfig:data_flow_schematic}, that begins with randomly generating several thousand spherical Au nanoparticles, placed atop unique amorphous C substrates, with random radii, orientations, locations, and with possible twin defects or stacking faults, to account for structural diversity in the target micrographs. We utilize Prismatic \cite{rangel_dacosta_prismatic_2021, pryor2017streaming, ophus2017fast} to simulate HRTEM images using the multislice algorithm \cite{cowley1957scattering} and calculate their corresponding, ground-truth supervision labels, i.e., the sets of pixels in the images where nanoparticles are located. For each structure, we simulate HRTEM output waves at 300kV with a final resolution of 0.02 nm/pixel. From each simulation, multiple images of each nanoparticle structure are sampled under varying image conditions and noise. In order to facilitate targeted data curation when training neural networks, we extensively track and aggregate metadata at each phase, so that specific distributions of simulated data can be easily drawn from the full database.

\subsection{Image Segmentation of Nanoparticles with Supervised Neural Networks}

Utilizing our data generation pipeline, we examine how aspects of the training set of simulated HRTEM images, as induced via data curation, affect neural network segmentation performance on experimental HRTEM images. By evaluating both general trends and more granular effects of dataset characteristics on neural network performance, we identify data curation strategies for training high-performance ML models for experimental HRTEM image segmentation using only simulated data. To isolate the effects on model performance due to characteristics of the training dataset, we fix our neural network architecture and optimization hyperparameters and train multiple UNet networks \cite{ronneberger_u-net_2015} for each data condition. For a given data condition, training data are drawn I.I.D. from the simulation database, such that each network has both unique random initializations of learnable weights and independent data streams. In order to understand our model performance in the context of the intrinsic variability of HRTEM data, we benchmark our neural network performance against three previously published atomic-resolution experimental datasets \cite{groschner_machine_2021, sytwu_understanding_2022}, taken on two different aberration-corrected TEMs by different operators at ultra-high magnification (about 0.02 nm/pixel), because contrast mechanisms in HRTEM are highly dependent on sample thickness, chemical composition, structure, and experimental conditions. The first dataset comprises images of large Au nanoparticles (5nm) and agglomerates \cite{groschner_data_2020}; the second small Au nanoparticles (2.2nm)\cite{sytwu_data_2022}; and the final small CdSe nanoparticles (2nm)\cite{groschner_data_2020}. Simultaneous benchmarking on several different experimental datasets provides an opportunity to analyze how robust our trained models are to distributional shift and to determine the generality of training data effects. Previous best results on these datasets, when analyzed with neural networks trained with experimental data, are F1 scores of 0.89, 0.75,and 0.59, for the large Au \cite{groschner_machine_2021}, small Au \cite{sytwu_understanding_2022}, and CdSe \cite{groschner_machine_2021} datasets, respectively.

\begin{table}[hbtp]
    \centering
    \footnotesize
    \begin{tabular}{lR{1.75cm}R{1.75cm}R{1.75cm}R{1.75cm}R{1cm}R{1cm}}
    \toprule
     \multicolumn{2}{c}{Training Dataset} &
     \multicolumn{3}{c}{F1-score on Exp. Data} &
     \multicolumn{2}{c}{$N_\text{Epochs}$ to F1} \\
     \cmidrule(lr){1-2}\cmidrule(lr){3-5}\cmidrule(lr){6-7}
     Dataset & $N_{\text{Images}}$ & 5nm Au & 2.2nm Au & 2nm CdSe & $V_{90}$ & $V_{95}$ \\
    \hline
         Baseline & 512 &  0.710 &    0.740 &    0.681 &  19 &  21 \\
      Thermal &      512 &  0.727 &    0.767 &    0.621 &  20 &  23 \\
         All simulation effects &      512 &  0.822 &    0.814 &    0.647 &  21 &  24 \\
            Smaller NPs &  1024 & 0.833 &    0.809 &    0.673 &  11 &  13 \\
        Varying Substrate &     1024 &  0.885 &    0.842 &    0.620 &  11 &  12 \\
    Optimized Au &      8000 &  0.915 &    0.808 &    0.648 &   3 &   3 \\
    Optimized mixed Au/CdSe & 8000 &  0.884 &    0.863 &    0.731 &  1 &  2 \\
    Optimized CdSe &      8000 &  0.799 &    0.852 &    0.752 &   2 &  24 \\
    \bottomrule
    \end{tabular}
    \normalsize
    \caption{Best performance from neural networks on segmentation of nanoparticles in HRTEM images after training on various sets of simulated data, as measured on three experimental datasets. In the rightmost columns, we record the median number of epochs taken for networks to reach validation F1 scores of 0.90 and 0.95 on simulated data during training.}
    \label{tab:perf}
\end{table}

\begin{figure}[hbtp]
    \centering
    \includegraphics[width=6in]{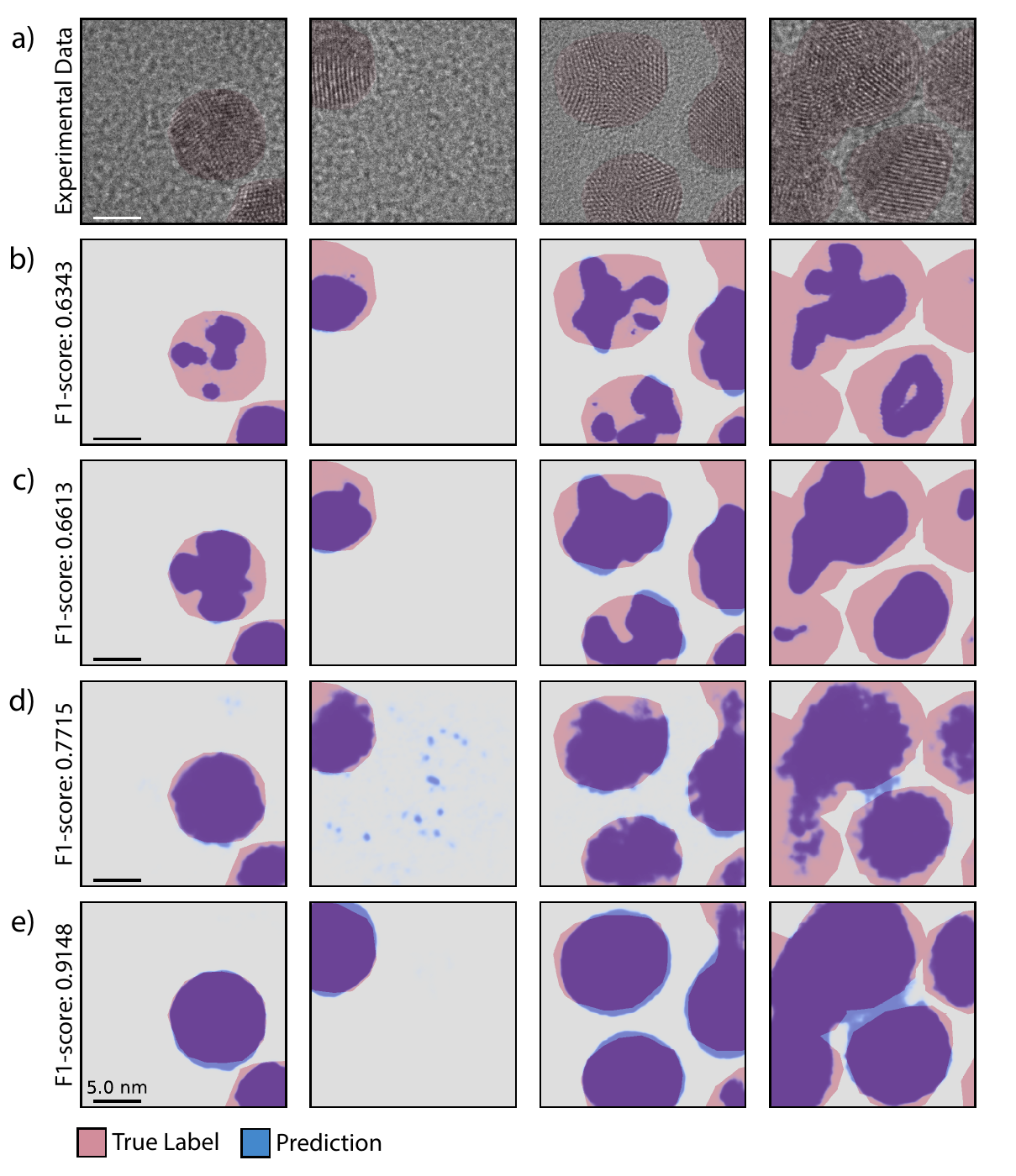}
    \caption{Characteristic performance of neural network models analyzing experimental images of Au nanoparticles after training on only simulated data. Models were selected trained on Baseline, Substrate, and Optimized Au datasets, from which these models represent median to strongly-performing examples. Scalebar is 2.5nm.} 
    \label{fig:segmentation_viz}
\end{figure}

Neural networks can be trained to segment experimental HRTEM datasets with moderate accuracy even with small simulated datasets containing only 512-1024 images, as indicated by our results in Table \ref{tab:perf}. Models trained on simulated data optimize performance on the simulated training data quickly and to an extremely high degree of accuracy---frequently achieving F1-scores above 0.90 on the (simulated) validation dataset in a small number of epochs (c.f. Table \ref{tab:perf}, rightmost column)---whereas their performance on experimental data increases more slowly, continually improving throughout training, even after performance on simulated data has apparently saturated (c.f. Fig. \ref{sfig:gen_dynamics}). Thus, performance on simulated data is not a reliable signal for performance on experimental data and benchmarking is crucial for successful model deployment. Model performance on validation data can lag behind training performance, reducing with increasing dataset size, similar to delays observed for models trained on small, algorithmically generated datasets \cite{power_grokking_2022}. In Fig. \ref{fig:segmentation_viz}, we visualize characteristic segmentation performance of four neural networks, across a range of accuracies, on the large Au dataset. Poorly performing segmentation models (Fig. \ref{fig:segmentation_viz}b, c), after training on just simulated data, can accurately predict segmentation regions on nanoparticles with clear lattice fringes, but might miss similar particles in other micrographs and/or lose significant performance when predicting segmentation regions for more complex structures textures and particles with many grains. Better networks have smoother, more consistent predictions (Fig. \ref{fig:segmentation_viz}d, e) but still might miss regions in particles with more complex structures or might have high-frequency spatial fluctuations in their predicted regions (Fig. \ref{fig:segmentation_viz}d, rightmost column), which are not physically consistent with nanoparticle structures, potentially indicating important noise features or aspects of the imaging conditions are not fully captured during data curation. Across the board, neural networks seem to segment nanoparticles more consistently when the particle (or particle grain) has visible lattice fringes, indicating that trained models can distinguish ordered lattice textures from other regions. 

\begin{figure}[hbtp]
    \centering
    \includegraphics[width=2in]{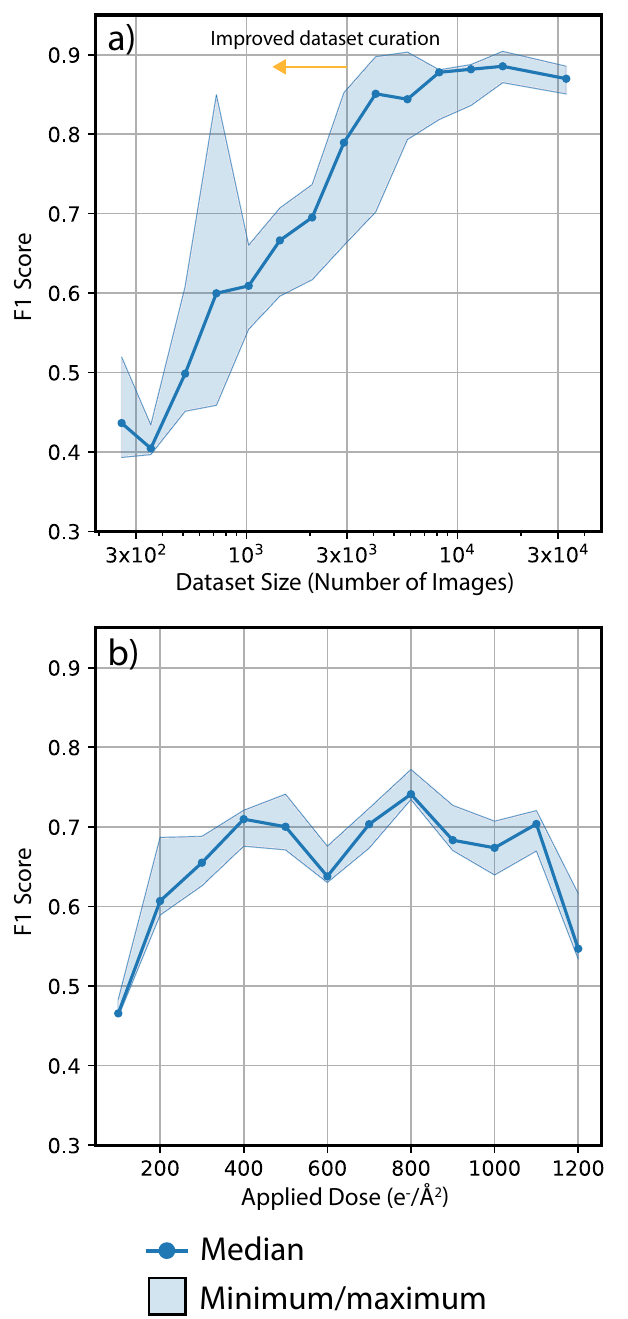}
    \caption{Effect of training dataset size (a) and noise from applied electron dose (b) on neural network performance, as measured on experimental data of large Au nanoparticles. Networks in top panel were trained on a uniformly random chosen subset of all simulated data in this study. The gold circle indicates the performance of the best neural network trained on a dataset containing only 1024 images, which were drawn from a simulated dataset comprising structures with varying substrate thicknesses. For each dataset condition, five randomly initialized neural networks were trained and measured.}
    \label{fig:data_size_noise}
\end{figure}

The overall cost of the model development process depends both on the cost of procuring effective training datasets and the cost of training the models themselves, and often, compromises must be made between acceptable costs and the end quality of obtained models. As is well known, neural networks trained with supervised learning methods can be greatly improved by using larger training datasets, which can cost much more to obtain and can cause model training time to increase substantially. In Fig. \ref{fig:data_size_noise}a, we measure the performance of neural networks on the large Au experimental dataset after training on datasets of increasing size, where the training data are drawn uniformly randomly from all of the aggregated databases (see Supplemental section \ref{si_data}). Segmentation performance saturates at an F1-score of about 0.9 after about 8000 images are included in the training dataset, after which there are only marginal improvements due to dataset size alone. With smaller datasets, model performance can be highly variable, and the gap between the worst and best model trained steadily decreases as the dataset size increases. By tuning the quality and composition of the simulated data, we can train comparably accurate models with more data-efficient curation strategies, as, for example, with the networks trained on structures with varying substrate thicknesses (c.f. Tables \ref{tab:perf} and \ref{stab:all_perf}), indicated by the gold circle in Fig. \ref{fig:data_size_noise}a. 

At a fixed dataset size, the model performance can be highly sensitive and dependent on the noise of the training dataset, as shown for models trained on datasets varying only in applied electron dose in Fig. \ref{fig:data_size_noise}b. On other experimental datasets, we find that model performance maintains this sensitivity, but now peaks at lower doseage rates, with slow fall off of performance as noise decreases (c.f. Figs. \ref{sfig:small_Au}b and \ref{sfig:cdse}b), indicating that noisier data are important for models to learn. Differences in the qualitative performance trend across noise levels between the large Au data and the other experimental data might be a result of differing noise distributions in the images arising from differing camera statistics. Given the sensitivity of model performance to the noise level of the training dataset, we recommend to sample relevant noise from wider distributions, such that during training models see examples from a range of signal-to-noise conditions, which can improve the consistency of model training but still requires a careful choice of the noise distribution (c.f. Fig. \ref{sfig:rng}). Our results appear to be consistent as to prior work for a similar nanoparticle segmentation task \cite{larsen_quantifying_2023} with a different experimental geometry in which nanoparticles are mounted on a crystalline substrate and are imaged over vacuum. Given that, in our task, the imaging beam passes through both the nanoparticle and the (amorphous) substrate and has higher effective electron dose (200--600e-/$\text{\AA}^2$ across all datasets), it is likely that the relationship to noise could be more complex and that the minimal experimental dose that could be consistently segmented is larger, i.e., more signal-to-noise is required for our task.

\begin{figure}[hbtp]
    \centering
    \includegraphics[width=6in]{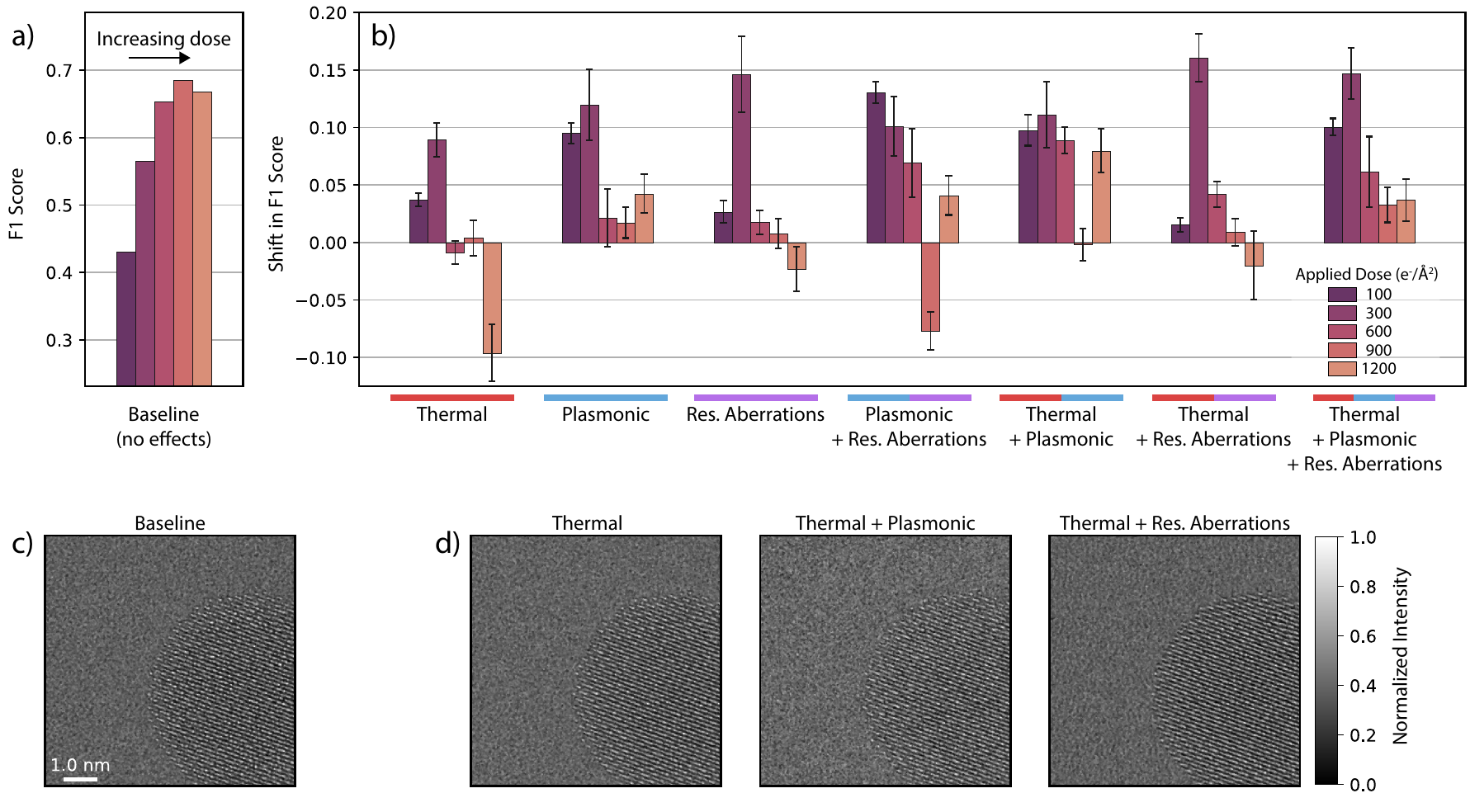}
    \caption{Baseline neural network performance with cheapest simulation methods (a) and effect of improving simulation fidelity on network performance, conditional on applied dosage, including effects from thermal averaging, residual aberrations, and plasmonic losses (b). Visual examples of baseline quality (c) and applied simulation effects (d) on simulated image of Au nanoparticles at 300kV with focal spread and 400$e^-/\text{\AA}^2$.}
    \label{fig:sim_effects}
\end{figure}

In high-throughput simulation settings, instead of producing many examples over varying noise conditions, individual examples in the training dataset could be made more informative by improving the quality of the TEM simulation, which can improve the robustness of the desired dataset and shift the source of development costs. In our case, improving the fidelity of HRTEM image formation by including the effects of inelastic scattering due to thermal vibrations, losses due to plasmonic excitations, and/or residual aberrations in the optical alignment of the microscope has a significant, positive effect on neural network performance. In Fig. \ref{fig:sim_effects}, we show the shift in performance of neural networks in the data scarce regime (N=512) when trained on data with combinations of applied thermal effects, residual aberrations, and plasmonic losses as compared to a baseline dataset with no such effects. Visually, the impact of these effects can be subtle on the simulated image (Fig. \ref{fig:sim_effects}c,d), yet, when added to the simulation data used for training, model performance can increase by as much as 0.1--0.15 in F1-score (Fig. \ref{fig:sim_effects}b). Importantly, in a regime of stable optimization, including simulation effects appears to be helpful across a wide range of applied dosage only when all additional effects are included. Of these effects, applying thermal effects is the most computationally expensive, as it requires averaging several HRTEM wavefunctions over a set of independent frozen phonon configurations, which linearly increases simulation costs, postprocessing time, and memory requirements, though, typically, only a small number of frozen phonons (O(10)) are needed to thermally converge TEM simulations of larger atomic structures. Applying the effects of residual aberrations and plasmonic losses are both relatively cheap in comparison, and thus, should be included in training datasets if possible.

\begin{figure}[hbtp]
    \centering
    \includegraphics[width=6in]{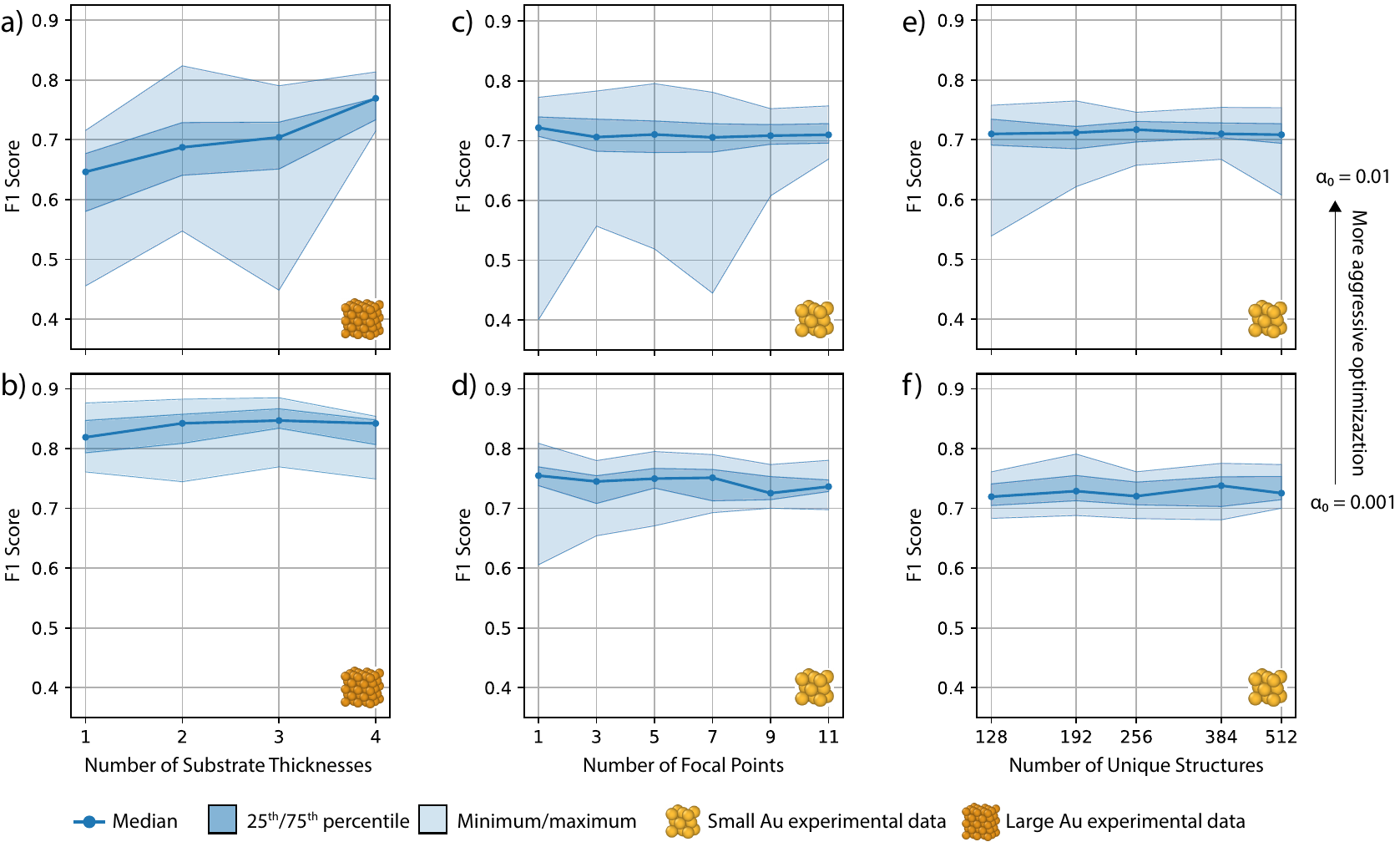}
    \caption{Effect of training dataset composition on neural network performance, at fixed noise levels and dataset size (N=1024 images), varying (a,b) the number of substrate thicknesses of the carbon substrate in the simulated structures, (c,d) the number of unique focal conditions in the training dataset, and (e,f) the number of unique structures in the dataset.}
    \label{fig:data_composition}
\end{figure}

The relationship between dataset composition---such as diversity of atomic structures or imaging conditions---to model performance is more nuanced, and in particular, aspects of dataset composition seem to be more important for controlling the variance of model performance than for boosting performance ceilings. That is, by including images of nanoparticles from a wide variety of structures or imaging conditions, we can much more likely guarantee that a randomly initialized model will optimize well and learn to segment experimental data effectively. In Fig. \ref{fig:data_composition}, we measure the performance of models trained on simulated datasets of fixed size (N=1024) with varying atomic structural content and imaging conditions. Networks trained on datasets comprising images of samples with varying substrate thicknesses performed better and with lower variance than networks trained on a single, fixed substrate thickness (Fig. \ref{fig:data_composition}a, b), indicating diversity in the atomic structures seen during training is important for image segmentation. Regarding imaging conditions, relative defocus is known to have a strong nonlinear effect on contrast features in HRTEM, which is reflected in a highly variable nonlinear relationship of model performance to the focal point of simulated data (c.f. Fig. \ref{sfig:defous_sensitivity}). To address this, we can sample images from a larger variety of focal points (Fig \ref{fig:data_composition} c,d), increasing the number of unique focal points (and by construction, focal range) seen during training while maintaining the total size of the dataset, which has a small positive effect on segmentation performance and a strong effect on reducing model variance. The number of unique atomic structures, however, does not significantly impact model performance (Fig. \ref{fig:data_composition}e,f), as long as a wide variety of imaging conditions are indeed being sampled in the training dataset. In general, these data compositional effects are more significant when models are optimized more aggressively (Fig. \ref{fig:data_composition}a,c,e) by using a larger initial learning rate and the extra data diversity can be interpreted as having a regularizing effect. Notably, regularization effects arising from an increase in data diversity might come into play only when the training dataset is already a suitable match to the experimental data being analyzed---i.e., that the distributions from which simulated training data and experimental data are drawn are similar. We find that these regularization effects are not noticeable when neural network performance is measured on the large Au or CdSe datasets (Figs. \ref{sfig:large_Au} and \ref{sfig:cdse}), where performance is worse overall, possibly due to the experiments measuring nanoparticle structures that are structurally dissimilar to the simulated data used to train the networks varying in focal conditions. Changing the nanoparticle size in the training data can have a strong effect on performance (c.f. Fig. \ref{sfig:gen_structure}), but does not appear to be completely related to the size of the nanoparticles in experimental images.

Incorporating these lessons on data effects altogether, we can design an optimized simulated dataset to target maximal absolute performance on our experimental benchmarks---ultimately, the balance of dataset size, fidelity, and composition will be dictated by the needs of a particular experiment or analytical task, which can include constraints on model size, data curation time, and model time-to-train. Here, we focused on improving performance on both the large Au dataset and the CdSe dataset by increasing the variety of atomic structures seen, including all simulation effects mentioned previously, sampling a wider range of imaging and noise conditions, and using a modestly large training dataset (8000 total images). On these benchmarks, with slight re-tuning of the training hyperparameters, we achieve a maximum F1 score of 0.9189 and 0.7516, for Au and CdSe, respectively, whilst also achieving relatively strong generalization performance (c.f. Fig. \ref{sfig:gen_opt}). When using a random 50/50 mixture of the optimized datasets, networks have strong generalization performance across the three datasets, but better peak performance is achieved only on the small Au dataset (c.f. Table \ref{tab:perf}). For details regarding the composition of these datasets and the training strategy and performance measurements on other auxiliary datasets, please refer to supplemental sections \ref{si_opt_au} and \ref{si_opt_cdse}.

\section{Discussion}

Our results, taken altogether, indicate that high-accuracy supervised models for analyzing atomic-resolution HRTEM experiments can be trained with sufficient high-quality simulated databases. These synthetic datasets tend to need to be larger than what is typically curated experimentally (around 4000-8000 images), simulated with relatively high fidelity, and must contain appropriate dispersions of atomic structures, especially varying substrates, imaging conditions, particularly defocus, and applied noise to ensure consistent model performance. Critically, many observed effects of data curation strategies can be unnoticeable when networks perform poorly across the board or when the simulated dataset differs significantly from an experimental benchmark, as demonstrated by our tests varying the dataset composition at fixed dataset sizes (Fig. \ref{fig:data_composition}. With precise control of the prior distributions from which training datasets were drawn, we were able to isolate these successful data curation strategies for machine learning model development, which otherwise can become almost intractable due to the large observed variance of model performances under certain training dataset conditions.

The advantages of simulated data primarily arise from access to arbitrarily large datasets, a broader and controlled distribution of training data, and physics-based ground-truth measurements on the dataset. Our results indicate that, even when dealing with experimental data that arises from a highly-complex measurement process, the use of simulated data during the development of ML models can compete and even provide distinct performance gains as compared to utilizing experimental data. Alternative approaches for bridging the gap between simulated data and experimental data could involve training a complementary sets of neural networks to specifically learn how to model and apply noise effects and features not captured by simulation models, e.g., with CycleGANs \cite{khan_using_2023}, though the extent to which such an approach could be as effective for HRTEM as for annular dark-field STEM is unclear. Well-curated experimental datasets are still immensely valuable---both as reference benchmarks, as used in this study, and as high quality training data. An important caveat of our results is that all performance metrics shown here ultimately compare physics-based ground-truth labels and expert manual labels. Given the inevitability of human measurement error, bias, and uncertainty, the two sets of labels never match perfectly, leading to a performance gap in our models and thus an artificial performance ceiling. Further work is needed to best understand measurement error in the face of such distribution gaps and to understand the best way to develop machine learning models that can make use of such mixed data given systematically different labels. 

Ideally, during training, a model would have access to a simulated dataset drawn from an identical distribution to the experimental prior and enough examples in this dataset to provide adequate coverage of the experimental distribution in which it will be deployed for analysis. In this setting, ML models only make in-distribution predictions, and never face extrapolation errors. In practice, it can be difficult to determine the exact experimental prior, suggesting instead a strategy of drawing training data from a distribution which the experimental distribution could be a feasible subset, thus, reducing adverse effects of distributional shifts during model deployment. General-purpose tools like Construction Zone, which enable fast sampling of realistic synthetic examples for ML problems and are not limited to any specific domain, are a crucial component of robust data workflows which would enable such sampling strategies. Further, such general purpose tools enable careful tuning of training datasets to best match experimental prior distributions, which can be performed manually or automatically, for example, as a component in an active learning training loop \cite{ang_active_2021, vandermause_fly_2020}. Curating the best suitable dataset for a particular ML problem is a crucial component of reliable ML workflows, but should be pursued in conjunction with finding the best model framework, which can include developing bespoke model architectures \cite{xie_crystal_2018, musaelian_learning_2023} or implementing custom loss functions \cite{kervadec_boundary_2021} with regularization \cite{kirkpatrick_pushing_2021} specific to the problem of interest.

In sum, we have developed Construction Zone, a new Python package that enables programmatic generation and sampling of atomic nanostructures, a general purpose tool designed primarily designed to help study and design ML workflows at scale for nanoscale materials science problems. In conjunction with HRTEM simulation, we systematically generate large structural and imaging databases for training ML models and have demonstrated their utility for automated atomic-scale characterization with HRTEM. By systematically generating large structural and imaging databases, we achieve state-of-the-art performance on such models, whilst also providing the ability to design and develop ML tools carefully through high levels of specific control on the data generation process.

\section{Detailed Methods}

\subsection{Construction Zone}
Nanoscale atomic structures can be designed in Construction Zone (CZ) using a combination of Generator, Volume, Transformation, and Scene objects. Generators, which supply atoms in space, specify atom positions and can generate atoms in both crystalline and non-crystalline arrangements. Volume objects define the convex regions in which atoms from Generators are accepted. Volumes can either be defined by sets of convex algebraic surfaces, convex hulls of point clouds, or intersections of supplied convex objects; non-convex geometries can be created as the union of multiple Volumes. The Transformation module manipulates objects with standard routines like symmetry operations and more complex routines like applying inhomogenous strain fields or modifying the local chemistry of a structure. Generators and Volumes can be manipulated jointly or individually for full construction flexibility. For example, a grain can be oriented either by rotating a Generator lattice relative to its boundary or by rotating the whole object in global coordinates. Finally, Scenes aggregate objects together. Given a set of atomic objects, a Scene will handle interactions through a generation precedence scheme and prepares data for interfacing with other simulation methods or for writing to file. 

CZ also features a small set of auxiliary modules for more complex functionality and convenience. The Surface module provides fast routines for analyzing, querying, and modifying the surface of a generated object using derivatives of alpha-shape algorithms \cite{stukowski_computational_2014, edelsbrunner_three-dimensional_1994}. The Utilities module provides several analysis routines, such as radial distribution function (RDF) analysis and orientation sampling. Lastly, the Prefab module contains routines that generate ``pre-packaged'' objects like Wulff constructions of nanoparticles (as implemented in \cite{rahm_wulffpack_2020}) or grain structures with planar defects. For a full discussion of underlying routines, available features, and package usage, we refer the reader to the CZ documentation\footnote{https://github.com/lerandc/construction-zone, https://construction-zone.readthedocs.io/en/latest/}.

\subsection{TEM simulation}

High-resolution TEM image simulations in this study were performed using the multislice algorithm \cite{cowley1957scattering}, as implemented in the Prismatic software package \cite{rangel_dacosta_prismatic_2021}. In the multislice algorithm, we model the image created in an electron microscope through the interaction of an electron beam and a material sample as the evolution of a 2D complex wavefunction, $\psi(\mathbf{r})$. The evolution of the wavefunction is given by the Schr\"odinger equation for fast electrons \cite{van1985advances}

\begin{equation}
\label{E:schrodinger}
    \frac{\partial \psi(\mathbf{r})}{\partial z} 
    = 
    \frac{\ii \lambda}{4 \pi} 
    \nabla^2 \psi(\mathbf{r})
    +
    \ii \sigma V(\mathbf{r}) \psi(\mathbf{r})\,,
\end{equation}

where $V(\mathbf{r})$ is the electrostatic potential of the sample and $\sigma$ is an interaction constant. $V(\mathbf{r})$ is typically calculated with an isolated atom approach, where the total potential is the sum of independent atomic potentials

\begin{equation}
    V(\mathbf{r}) = \sum_i V_i (\mathbf{r})
\end{equation}

where the individual atomic potentials themselves calculated with a parameterized look-up table of electron scattering factors \cite{kirkland_advanced_2010, lobato_accurate_2014}. Alternatively, the scattering potential can be determined for a sample through ab-initio techniques \cite{abtem}. Once $V(\mathbf{r})$ is determined, it is split into a series of binned slices along the beam direction and then the wavefunction $\psi(\mathbf{r})$ is evolved through a split-step method, where the electron beam alternately interacts with the sample

\begin{equation}
    \label{Eq:transmission}
    \psi(\mathbf{r})
    = T(\mathbf{r}) \psi_0(\mathbf{r}) =
    e^{\ii \sigma V_n(\mathbf{r})} \psi_0(\mathbf{r}),
\end{equation}

and then propagates in free space 

\begin{equation}
    \label{Eq:propagator}
    \psi(\mathbf{r})
    =
    \invFFT
    \left\{
    e^{\ii \lambda \Delta z |\mathbf{k}|^2} 
    \FFT\Big[
     \psi_0(\mathbf{r}) 
    \Big] \right\}.
\end{equation}

After the wavefunction has propagated through the entire sample, we obtain the exit wavefunction, which can be further modified to apply defocus and other residual optical aberrations typically seen in HRTEM imaging by applying another transmission operation

\begin{equation}
\label{eq:abb_wave}
    \Psi(\mathbf{k}) = \Psi_0 ((\mathbf{k}) \exp[-i \chi((\mathbf{k})]
\end{equation}

where $\Psi_0 ((\mathbf{k})$ is the unaberrated wavefunction in Fourier space and $\chi$ is the aberration function. 

\subsection{Database Generation with CZ and HRTEM simulation}

We simulate a HRTEM data stream by first generating a large database of semi-realistic structures and then simulating micrographs for those structures under suitable imaging conditions. We greatly approximate the variety of nanoparticles imaged in \cite{groschner_machine_2021} as a series of spherical Au nanoparticles of varying diameters with a small number of planar defects, which are equally likely to be twin defects or stacking faults. Each nanoparticle is placed upon a unique amorphous carbon substrate and at a random location on the substrate surface at a random orientation. We then perform plane-wave multislice simulations of the structures at an acceleration voltage of 300kV, focused to the center of the atomic model, at a resolution of 0.02nm/pixel. We simulate output wavefunctions for both the full structure and just the nanoparticle in vacuum. To create ground truth segmentation masks, we threshold the phase of the output wavefunctions from the nanoparticles in vacuum, averaged over frozen phonons. Threshold values were determined heuristically. We found phase-thresholding to be much more stable and physically consistent across focal conditions than analogous intensity thresholding techniques.

With the output wavefunctions, we then apply relevant focal conditions, residual aberration effects, focal spread, thermal effects, plasmonic losses, and electron dosage effects. Focal point, focal spread, and residual aberrations are applied to output wavefunctions using Eq. \ref{eq:abb_wave} with the appropriate aberration function; the effect of focal spread is approximated by incoherently averaging the wavefunction over a series of focal points distributed about the intended focal point. Thermal effects are included via the frozen-phonon approximation, where we incoherently average several exit wavefunctions to approximate the effect of thermal motion, corresponding to unique, perturbed copies of the original atomic structure. The atomic positions are perturbed by I.I.D. Gaussian noise, with amplitudes given by atomic Debye-Waller factors; for all simulations including thermal effects, we used eight frozen phonons. We approximate plasmonic losses in HRTEM imaging by applying a contrast reduction to the wavefunction intensities before averaging over frozen phonons. The contrast reduction is applied by taking a weighted pixelwise mean of the original intensity and an intensity with a constant added background (renormalized to have its original mean value). Noise from applied electron dosage is modeled by sampling count intensities from scaled Poisson distributions. We note that we do not include potential artifacts from the camera, i.e. via the modulation transfer function, though all experimental datasets have been acquired by scintillator-based detectors, for which including effects from the modulation transfer function in the noise sampling process could further improve performance \cite{larsen_quantifying_2023, madsen_deep_2018}. 

\subsection{Neural Network Training}

For our neural network development, we used a standard UNet architecture \cite{zhou_unet_2020, groschner_machine_2021, ronneberger_u-net_2015} with a ResNet 18 \cite{he_deep_2016} encoder/decoder architecture and three pooling/upsampling stages, resulting in about 14 million trainable parameters. Models were trained using a mixed categorical cross-entropy and F1-score loss function on the segmentation masks. We performed a brief manual hyperparameter search to determine initial learning rates and learning rate schedules which were suitable enough for stable optimization under a wide variety of the simulated datasets. For most networks whose performance is reported in this study, we used the Adam optimization algorithm \cite{kingma_adam_2017} with initial learning rates of 0.01 and 0.001, and a constant learning rate decay of 0.8. Unless explicitly indicated otherwise, results in the primary text reflect those of networks trained with an initial learning rate of 0.001. Models were trained for a fixed number of epochs (25) with no early stopping; the network parameters that achieved the lowest validation loss were saved via checkpointing. Input data were first normalized, per image, to a range of 0 to 1, and then augmented with orthogonal rotations and random left-right/up-down flips. For experimental data, images were processed with a 3x3 median filter and then normalized to a range of 0 to 1, per image, before evaluating neural network test performance. Experimental image patches containing only substrate regions were removed from the training dataset prior to preprocessing. Per simulated training data condition tested, at least five different networks were trained with different random weight initializations and training data subsets (drawn uniformly at random after appropriate filtering). For each network, relevant metadata were saved alongside the training history and model parameters. All training was performed using Tensorflow version 2 \cite{abadi_tensorflow_2016}, with the Segmentation Models package \cite{yakubovskiy_2019}. Neural networks were benchmarked for performance on 3-5 experimental datasets using the F1-score

\begin{equation*}
    F_1 = \frac{TP}{TP + 0.5(FN + FP)}
\end{equation*}

which is a ratio of true positives (identifying nanoparticle region correctly) to true positives, false positives (identifying substrate as nanoparticle) and false negatives (identifying nanoparticle as substrate). The F1 score measures 0 when all nanoparticle pixels are misclassified and 1 only when there are zero misclassified pixels. 

\section{Data Availability}

Code used to generate atomic structures, HRTEM simulations, and fully processed images, as well as code used to train neural networks models, will be made fully available upon publication. Given the large amount of training data generated for this study, a small selection will be made fully available upon publication; other selections of training data will be available upon request.

\section{Author Contributions}

LRD performed the simulated data generation, machine learning, and data analysis and prepared the manuscript. KS and CKG acquired and labeled the experimental data, provided assistance with the machine learning, and helped prepare the manuscript; CKG also developed initial workflows for the machine learning. MCS led and supervised the study and helped prepare the manuscript.

\section{Acknowledgements}
The authors thank Dr. Colin Ophus and Dr. Philipp Pelz for helpful discussions. This material is based upon work supported by the U.S. Department of Energy, Office of Science, Office of Advanced Scientific Computing Research, Department of Energy Computational Science Graduate Fellowship under Award Number DE-SC0021110. K.S. was supported by an appointment to the Intelligence Community Postdoctoral Research Fellowship Program at Lawrence Berkeley National Laboratory administered by Oak Ridge Institute for Science and Education (ORISE) through an interagency agreement between the U.S. Department of Energy and the Office of the Director of National Intelligence (ODNI). Work at the Molecular Foundry was supported by the Office of Science, Office of Basic Energy Sciences, of the U.S. Department of Energy under Contract No. DE-AC02-05CH11231.

This report was prepared as an account of work sponsored by an agency of the United States Government. Neither the United States Government nor any agency thereof, nor any of their employees, makes any warranty, express or implied, or assumes any legal liability or responsibility for the accuracy, completeness, or usefulness of any information, apparatus, product, or process disclosed, or represents that its use would not infringe privately owned rights. Reference herein to any specific commercial product, process, or service by trade name, trademark, manufacturer, or otherwise does not necessarily constitute or imply its endorsement, recommendation, or favoring by the United States Government or any agency thereof. The views and opinions of authors expressed herein do not necessarily state or reflect those of the United States Government or any agency thereof.

\newpage
\section{References}
\bibliography{refs}

\end{document}


\title{Supplementary Information: A robust synthetic data generation framework for machine learning in High-Resolution Transmission Electron Microscopy (HRTEM)}

\author{Luis Rangel DaCosta}
\email{luisrd@berkeley.edu}
\affiliation{Department of Materials Science and Engineering, University of California Berkeley, Berkeley, CA 94720}
\affiliation{National Center for Electron Microscopy, Molecular Foundry, Lawrence Berkeley National Laboratory, 1 Cyclotron Road, Berkeley, CA, USA}

\author{Katherine Sytwu}
\affiliation{National Center for Electron Microscopy, Molecular Foundry, Lawrence Berkeley National Laboratory, 1 Cyclotron Road, Berkeley, CA, USA}

\author{C.K. Groschner}
\affiliation{Department of Materials Science and Engineering, University of California Berkeley, Berkeley, CA 94720}
\affiliation{National Center for Electron Microscopy, Molecular Foundry, Lawrence Berkeley National Laboratory, 1 Cyclotron Road, Berkeley, CA, USA}

\author{M.C. Scott}
\email{mary.scott@berkeley.edu}
\affiliation{Department of Materials Science and Engineering, University of California Berkeley, Berkeley, CA 94720}
\affiliation{National Center for Electron Microscopy, Molecular Foundry, Lawrence Berkeley National Laboratory, 1 Cyclotron Road, Berkeley, CA, USA}

\date{\today}
\maketitle

\section{Description of Data}\label{si_data}

\subsection{Simulated Datasets}

\subsubsection{Baseline}

Our initial baseline reference dataset contains 512 unique structures of FCC gold nanoparticles on amorphous carbon substrates, with 2048 simulations and 10240 training images. The nanoparticles are perfectly spherical and have a normally distributed radius between 3nm and 5nm (clipped range) with a mean of 4nm and a standard deviation of 1nm and have a lattice parameter of 4.08265 $\text{\AA}$. Each nanoparticle contains between 0 and 3 planar defects, chosen uniformly at random, which are equally likely to be either twin defects or stacking faults, and are at least three (111) planes from each other and the edge of the particle. The nanoparticles are placed onto an 12.8 nm x 12.8 nm x 5nm amorphous carbon substrates with a uniformly random distribution of atoms and a minimum bond length of 1.4$\text{\AA}$ \cite{ricolleau_random_2013}; each substrate is uniquely generated and obeys periodic boundary conditions in the lateral dimensions with respect to the carbon distribution. When the nanoparticle is placed onto the substrate, it is placed at a random orientation (uniformly distributed on the unit sphere) and at a distance of at least 1nm from the substrate edges. The nanoparticles are embedded slightly into the substrate, at a fraction of 0.2 of their radii. For each structure, we simulate 4 different HRTEM output waves at 300kV without applied thermal effects using the Prismatic software package \cite{rangel_dacosta_prismatic_2021} at different defoci, sampled from a normal distribution with a mean of 3nm and standard deviation of 2nm (in C1 convention, measured from the center of the simulation cell); the simulations were run with infinitely projected atomic potentials sampled at 0.1 $\text{\AA}$. For each simulation, we then apply Poisson noise at dose rates of 100, 300, 600, 900, and 1200 electrons/$\text{\AA}^2$ and then filter the image with a median filter with 3x3 pixel window. Finally, each training image is randomly cropped from its 640x640 pixel resolution to a 512x512 resolution. We run an equivalent simulation for each structure without the substrate in order to generate an automated segmentation label. For these label simulations, we threshold the intensity image of the nanoparticle in vacuum and then apply a Gaussian blur with a standard deviation of 5 pixels, to smooth the mask.\\

\subsubsection{Thermal}

Our high quality reference dataset, having the same dataset composition of structures, simulations, and images, is largely similar to the previous baseline dataset barring two major differences; it reuses the same underlying dataset of structures. First, each simulation is instead performed with 3D potentials and a slice subsampling of 32 and with 8 frozen phonon configurations. Second, to each simulation, we apply a defocus spread of 15$\text{\AA}$ by incoherently averaging the output wavefunction about its desired focal point with a series of 25 defocused wavefunctions up to $\pm$45 $\text{\AA}$ with normally distributed weights. Each simulation was saved with the complex outputwave of all frozen phonon configurations. Two versions of this database were generated, one with thermally averaged output images and one with post-processed images utilizing only a single frozen phonon configuration. For generation of the relevant segmentation labels, we used instead a phase threshold on the output wavefunction, averaged over the frozen phonons. At these focal points, differences in the intensity and phase thresholding techniques are minimal; for the rest of the simulated data, which vary much greater in optical imaging conditions, we used the phase thresholding method, as it is more stable over imaging conditions as compared to the intensity threshold.\\

\subsubsection{Dose variation}

Our dose variation dataset is an addendum to the thermal dataset, containing further samplings of applied Poisson noise for the same structures and simulations. An additional 14,336 images are contained in this dataset, corresponding to seven sets of 2048 images each with dose rates of 200, 400, 500, 700, 800, 1000, and 1100 electrons$/\text{\AA}^2$. \\

\subsubsection{Simulation effects}

The plasmonic effect dataset is equivalent in composition to the Thermal dataset and uses the same structures and corresponding simulations. In addition to the thermal averaging across phonon effects, we apply the effects of plasmonic losses and residual effects in HRTEM imaging, combining each effect with each other in pairwise combinations, as well as altogether. Plasmonic losses are modeled by applying a contrast reduction to the wavefunction intensities before averaging over frozen phonons. The contrast reduction is applied by taking a weighted pixelwise mean of the original intensity and an intensity with a constant added background (renormalized to have its original mean value); for this dataset, we used a weight of 0.5 and an additive background of 1.0. For the residual aberrations, we apply a set of experimentally measured residual aberrations, as measured during an experimental session on the spherical-aberration corrected TEAM 0.5 microscope; these are included in the aberration function used to apply the baseline image defocus. We list the set of residual aberrations used in Table \ref{s_table:res_abs}, using the coefficient notation convention in \cite{ophus_automatic_2016, rangel_dacosta_prismatic_2021}.\\

\begin{table}[hbtp]
    \centering
    \begin{tabular}{lrr}
        \hline
         $C_{m,n}$ &  Magnitude & Rotation\\
         \hline
         $C_{2,2}$ &  1.602 nm& -68.5$^\circ$\\
         $C_{3,1}$ &  23.9 nm & 29.2$^\circ$\\
         $C_{3,3}$ &  17.98 nm& -170.7$^\circ$\\
         $C_{4,0}$ &  -1.913 $\mu$m & 0.0$^\circ$\\
         $C_{4,2}$ &  42.64 nm & -34.5$^\circ$\\
         $C_{4,4}$ &  158.3 nm & 89.2$^\circ$\\
         $C_{5,1}$ &  1.678 $\mu$m & 104.5$^\circ$\\
         $C_{5,3}$ &  528.7 nm & -12.7$^\circ$\\
         $C_{5,5}$ &  3.117 $\mu$m & -6.3$^\circ$\\
         $C_{6,0}$ &  1.048 mm & 0.0$^\circ$\\
         $C_{6,6}$ &  4.821 $\mu$m & 60.2$^\circ$\\
         \hline
    \end{tabular}
    \caption{Values of aberration coefficients used to sample residual aberrations in simulated datasets, as measured after standard microscope calibration procedures on the TEAM 0.5 microscope operating at 300kV. Notation follows \cite{ophus_automatic_2016, rangel_dacosta_prismatic_2021}.}
    \label{s_table:res_abs}
\end{table}

\subsubsection{Varying Substrate}

The Varying Substrate dataset is a dataset containing 1024 structures, their corresponding simulations, and 11264 final processed training images. The simulations for this dataset contained 8 frozen phonon configurations and no 3D potential subsampling, and otherwise follow the same simulation settings as the Thermal dataset; all remaining datasets utilize this simulation technique. The nanoparticles are sampled in a structurally equivalent method to those in the baseline datasets. The substrates, instead of having a uniform thickness of 5nm, have uniformly distributed thicknesses between 2nm and 8nm. For each structure (and simulation), we generate images at 11 different focal points, between -50nm and +50nm in steps of 10nm, about which individual focal points are normally distributed with a standard deviation of 2nm. Each image has an applied dose rate of 400 electrons$/\text{\AA}^2$, an applied focal spread of 15 $\text{\AA}$, and has the same set of residual aberrations in Table \ref{s_table:res_abs}. Otherwise, the post-processing is performed equivalently to those images in the Thermal dataset.\\

\subsubsection{Smaller NPs}

\begin{figure}
    \centering
\includegraphics[width=6in]{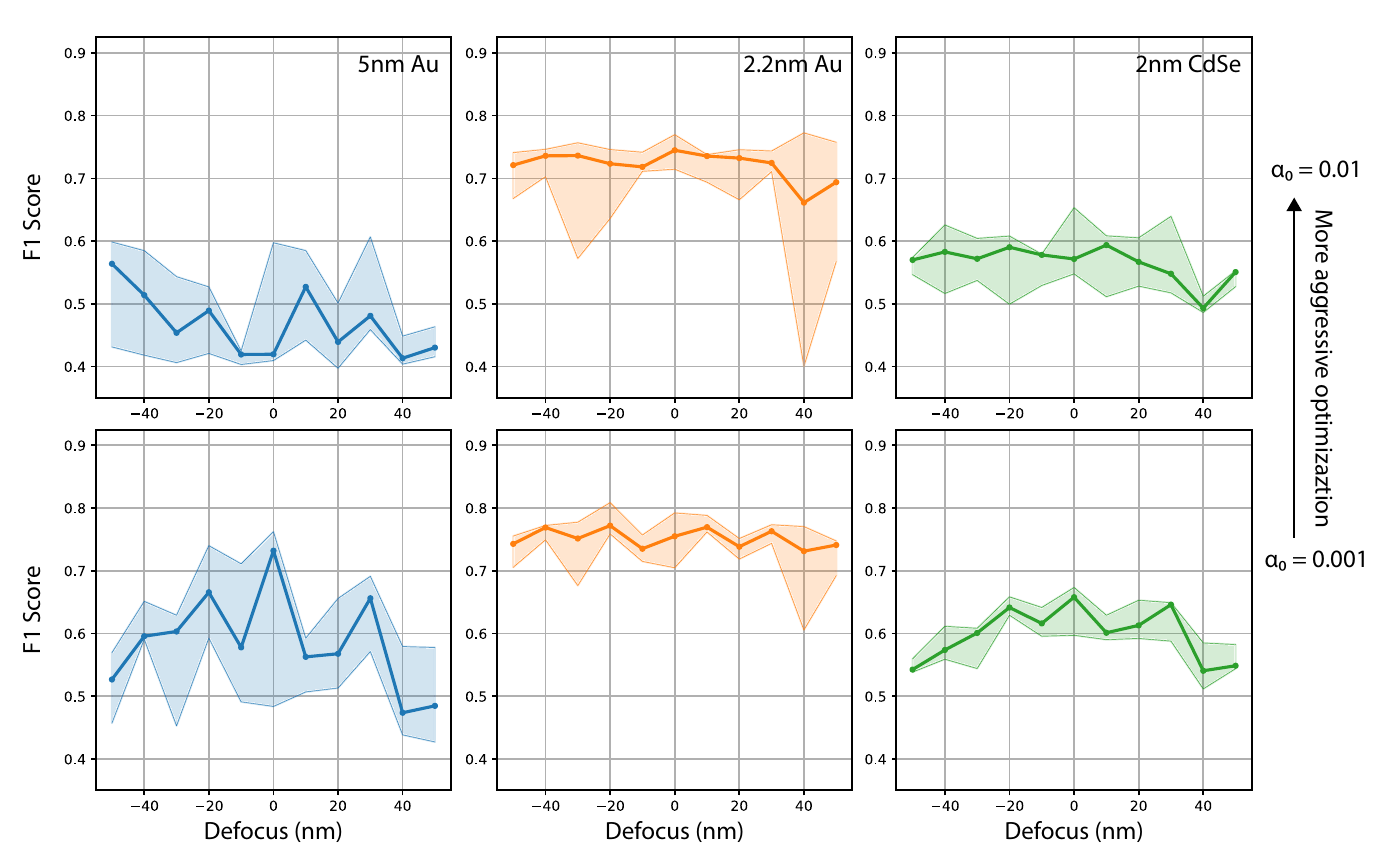}
    \caption{Effect of simulated focal point on neural network performance, as measured on three different datasets. Networks are trained on two different initial learning rates.}
    \label{sfig:defous_sensitivity}
\end{figure}

The Smaller NPs dataset is a dataset containing 512 structures, their corresponding simulations, and 11264 final processed training images. The nanoparticles in this dataset are sized in two distributions---the first 256 nanoparticles have a normally distributed mean radius of 1nm, with a standard deviation of 0.25nm, and are bounded by radii of 0.75nm and 1.25nm; and the final 256 nanoparticles have a mean radius of 2.5nm with a standard deviation of 0.75 nm, and are bounded by radii of 1.5nm and 3.5nm. The smaller group of particles has only up to 1 planar defect per particle, while the larger group can have up to 3 planar defects per particle. The carbon substrate is sampled equivalently to those in the baseline datasets. The simulation and post-processing of data is equivalent to those images in the Substrate Varying dataset, except that each structure is sampled twice per focal point. These doubled samples are used only when sampling data from a single focal point, i.e., the first point in Figure \ref{fig:data_composition}c and d, or for the data presented in Figure \ref{sfig:defous_sensitivity}.

\subsubsection{Optimal Au}\label{si_opt_au}

Our optimal Au dataset is a dataset designed to target high segmentation performance on the large Au experimental dataset from \cite{groschner_machine_2021} with reasonable data preparation, data storage, and training costs. The dataset consists of 2000 nanoparticle on carbon substrate structures, and 8000 final processed images, with four images per structure. The nanoparticles are generated with a broader size distribution, using a radius bounded betweeen 1.5nm and 5.0nm, with a mean of 3.5nm and a standard deviation of 3nm. Three-fourths of the nanoparticles follow the same defect distribution as before, with up to 3 planar defects per particle, while the remaining particles instead have random grain boundaries, The grain boundaries are generated by splitting the particle at a (111) plane, and taking the grain after the split and either randomly rotating it along the corresponding [111] axis or by randomly rotating it to an arbitrary rotation on the unit sphere; these particles were equally likely to have either type of grain boundary. The carbon substrates for this dataset have a thickness which is uniformly distributed between 3nm and 7nm. For this dataset, we combine the effects of all post-processing techniques used previously, and further diversify the dataset by sampling from broader distributions of imaging conditions. The final post-processed images have uniformly random focal points between -40nm and 40nm, a focal spreads between 10 and 25 $\text{\AA}$, and electron doses between 200 and 700 electrons/$\text{\AA}^2$. For the plasmonic effects, we use a uniformly random weight between 0.3 and 0.85 with a constant additive background of 1.0. In order to sample a random set of aberrations with physical consistency, we sample aberrations about the aberration distribution centered on the values in Table \ref{s_table:res_abs} by randomly scaling the magnitudes of each aberration by a multiplicative factor normally distributed about 1.0 with a standard deviation of 0.25 and apply an additive rotation normally distributed about 0 radians with a standard deviation of $\pi/4$. Median filtering and phase thresholding is performed in the same manner as in all previous datasets.\\

\subsubsection{Optimal CdSe}\label{si_opt_cdse}

Our optimal CdSe dataset is a dataset designed to target high segmentation performance on the CdSe experimental dataset from \cite{groschner_machine_2021} with reasonable data preparation, data storage, and training costs. Similarly to the optimal Au dataset, this dataset contains 2000 nanoparticle on carbon substrate structures, and 8000 final processed images, with four images per structure. The CdSe nanoparticles have a Wurtzite structure, with lattice parameters $a=4.39 \text{\AA}$ and $c=7.17 \text{\AA}$ (c.f. \cite{jain_commentary_2013}, mp-1070) and are perfectly spherical with normally distributed radii with a mean of 1.75nm, standard deviation of 3nm, and which are bounded between 1nm and 2.5nm. These nanoparticles contain Type I instrinsic stacking faults \cite{lahnemann_luminescence_2014}, separated from each other and the bounds of the particle by at least four basal planes. Particles with a radius greater than 2.2nm can have up to 2 stacking faults, with a radius greater than 1.5nm can have up to 1 stacking fault, and with a radius less than or equal to 1.5nm can have no stacking faults. The CdSe nanoparticles, otherwise, are generated similarly to the Au nanoparticles, with uniformly random orientations over the unit sphere and with slight embedding into the substrate surface. Each full structure contains up to 5 CdSe nanoparticles, and separated from each other and the boundaries of the substrate by 1nm. The number of nanoparticles per scene is chosen uniformly at random, and the locations of the nanoparticles are chosen uniformly at random via rejection sampling with a limited number of attempts, and thus, the final distribution of nanoparticles per scene is not entirely uniform. Substrates for this dataset have a thickness uniformly distributed 3nm and 7nm. Post-processing for the optimal CdSe dataset is done completely equivalently to the optimal Au dataset with only one major difference: the dose range used here is instead uniformly distributed between 150 and 450 electrons/$\text{\AA}^2$. We note that for the CdSe nanoparticle mask generation, we used a lower heuristic value for the phase thresholding technique, since Cd and Se are weaker electron scatterers than Au.\\

\subsubsection{Optimal Mixed Au/CdSe}\label{si_opt_mix}

\begin{figure}[hptb]
    \centering
    \makebox[\textwidth]{\includegraphics{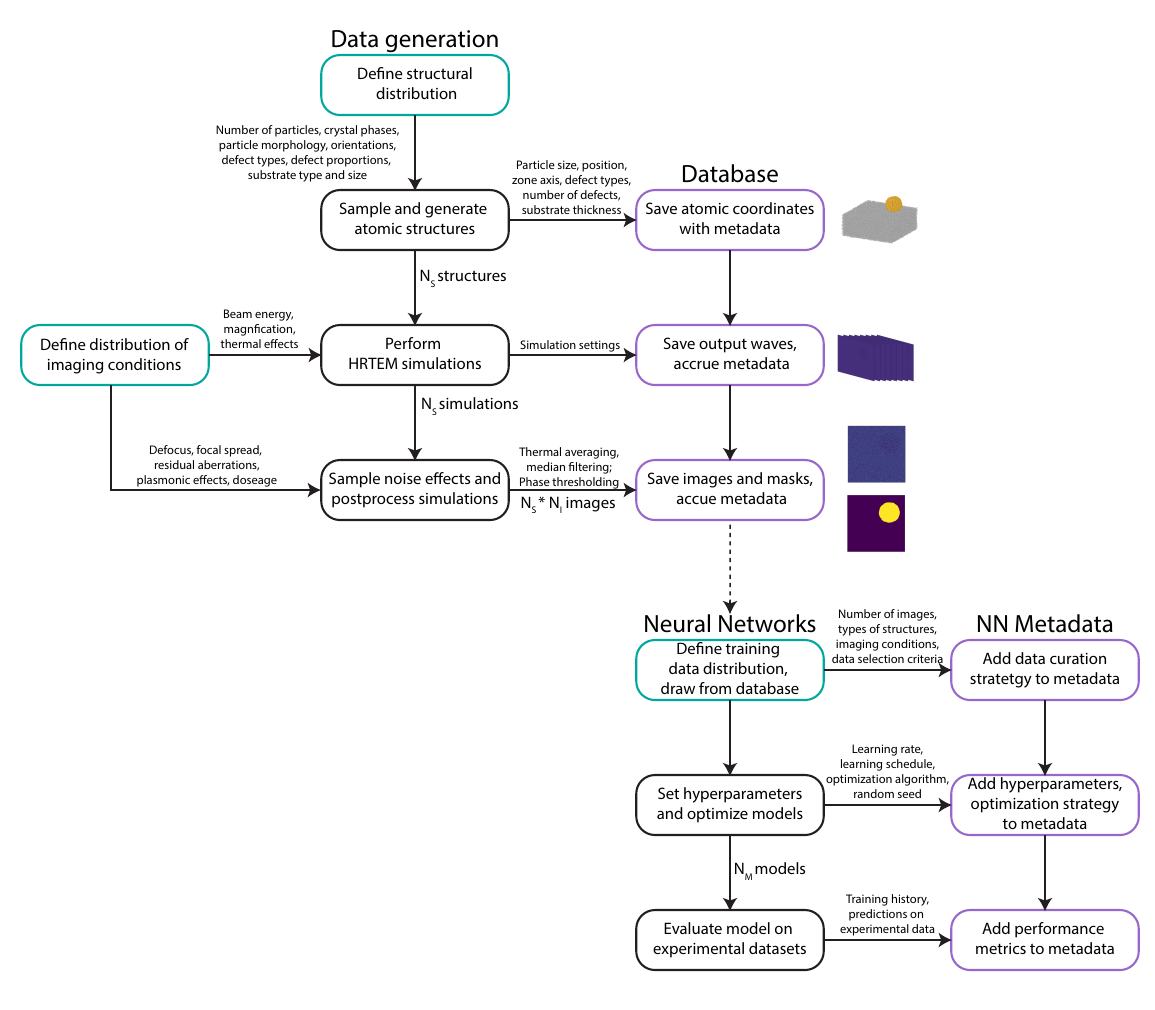}}
    \caption{Full end-to-end workflow schematic for database generation and neural network training as performed in this study. Information aggregated into metadata or operations performed to create data are indicated by the arrows linking steps in the workflow.}
    \label{sfig:data_flow_schematic}
\end{figure}

\subsection{Data sampling methodologies}

\subsubsection{Data selection during training}
For the vast majority of our training experiments, we trained five networks per data condition in order to reduce the effect of network training variance in our analysis. Each network received a uniformly random draw of samples from the underlying dataset. Three-fourths of the images were used to train the network, while the remaining fourth was used as a validation split. Images were further augmented during training with random orthogonal rotations and horizontal/vertical flips, though each epoch during training contained no more images than the number of unique, original examples in the training subset. For networks whose performance are tabulated in Fig. Fig. \ref{fig:sim_effects} and \ref{fig:data_size_noise}b, the dataset was subset by target dose and unique structures for a final sample of 512 images for each network. For networks whose performance was tabulated in the dataset size test (Fig. \ref{fig:data_size_noise}a) we drew a uniformly random sample of data from all available dataset sources, excluding the "Optimal" datasets used to target absolute performance. For networks whose performance was tabulated in Fig. \ref{fig:data_composition}a, we split the processed images into four thickness subsets (2-3.5nm, 3.5-5nm, 5-6.5nm, 6.5-8nm), and drew samples 1024 images from each possible combination of 1, 2, 3, and 4 subsets.

For the focal condition and focal vs. structure augmentation results (c.f. Fig. \ref{fig:data_composition}b and c), which were derived from networks trained on the small NP simulated database, we utilized a cross-validation sampling technique to improve the statistical power of our results. These networks were trained over dataset splits of 128, 192, 256, 384, and 512 unique structures and 3, 5, 7, 9, and 11 unique focal conditions. In order to remove the effect of number of images, we only used combinations of these subsets in which at least 1024 processed images were available. When the subset of images available was larger than 1024, we selected a sample of 1024 images uniformly at random from the available images.\\

Ideally, when measuring the effect of number of unique focal conditions on segmentation performance, we would oversample networks trained on datasets consisting of fewer focal conditions, in order for the performance measurements to have more comparable statistical power. For example, the 5 networks trained on data drawn from all 11 focal conditions would have an effective data distribution approximately matching that of the underlying processed dataset; here, we could make a relatively accurate statement about the network segmentation performance due to the data. Networks trained on data drawn from only 3 focal conditions, however, would have a much more variable underlying data distribution---it would take several hundred unique draws to have a similarly accurate approximation of the underlying data distribution, and the networks themselves would have an inherently high performance variability. Further, since HRTREM contrast is highly sensitive to focal conditions and since contrast mechanisms fundamentally differ in underfocused, overfocused, and in-focus conditions, we cannot guarantee through uniformly random draws that measured effect is derived from number of focal conditions and not just absolute focal condition range/coverage. \\

To combat these statistical challenges whilst limiting training expenses, we implemented two techniques to reduce the variance of the measured performance for datasets drawn from only 3 or 5 unique focal conditions: partitioning the focal conditions and performing slight oversampling the data subsets. For subsets of 3 unique focal conditions, we split the available focal conditions into partitions of underfocused focal points (-50nm through -20nm), "in-focus" focal points (-10nm through 10nm), and overfocused focal points (20nm through 50nm), such that a focal condition from each group was represented in the final draw of processed  images. For subsets of 5 focal conditions, we further partitioned the underfocused and overfocused subsets into groups of two focal conditions each. Finally, for a given number of unique structures, we sampled more draws of 3 and 5 focal conditions than for 7, 9, and 11 focal conditions. Since this test was slightly more variable than other training series we ran, we ran at least 20 data draws for each pair of structure/focal conditions, and 35 data draws for data draws using only 3 or 5 unique focal conditions. We note that, due to the partitioning, the data distributions are not equivalent due to the sampling process. A larger, more fair statistical test--perhaps using cross-validation techniques in the sample drawing and in the performance analysis--is recommended to more clearly delineate the effect of focal conditions (in number and diversity) on the effect of segmentation of HRTEM data. Our results are likely to be loosely general, and it is well within the realm of possibility that different trends arise in different imaging conditions with different structures being imaged.\\  

\subsubsection{Noise sampling}

\begin{figure}[htbp]
    \centering
    \includegraphics[width=6in]{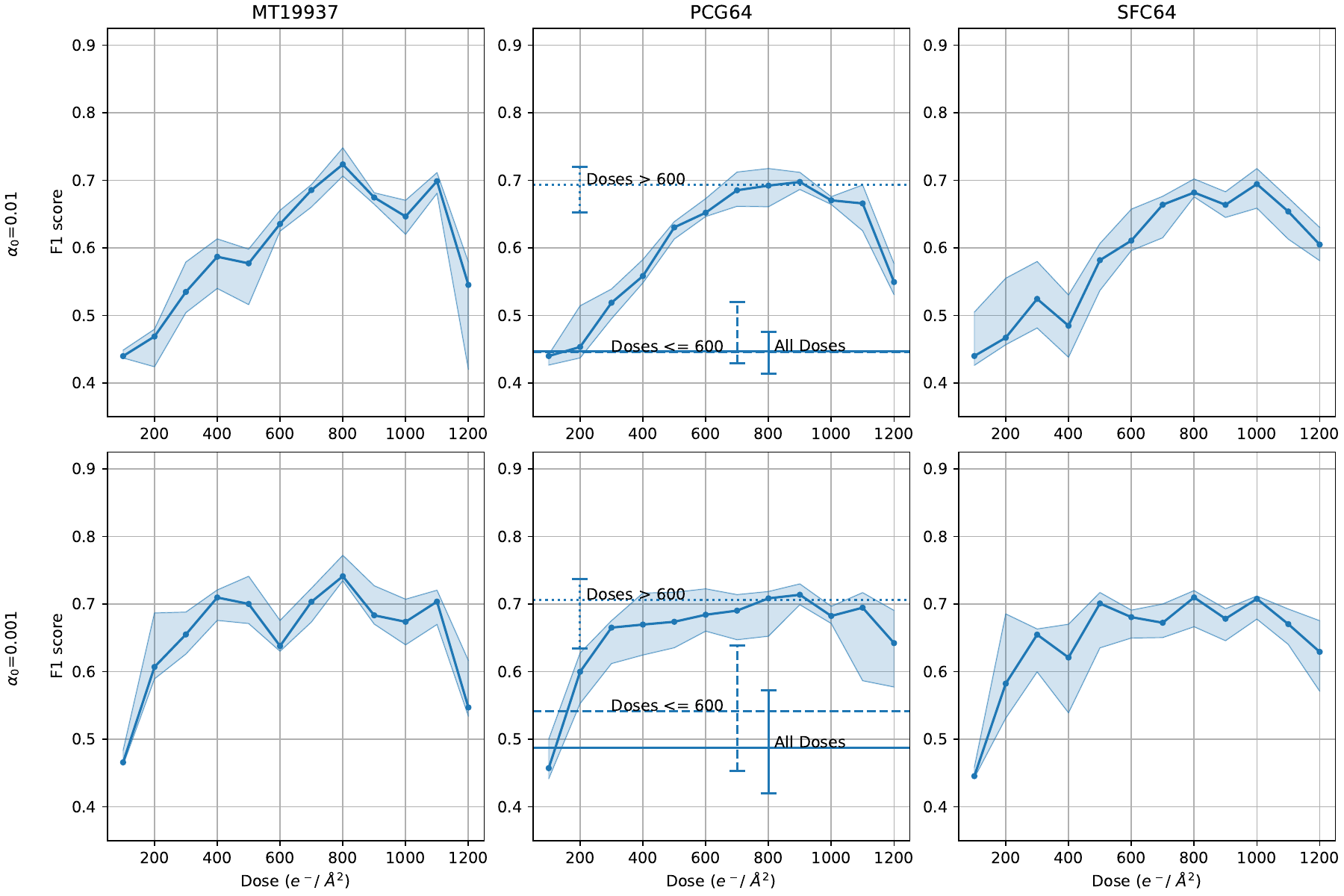}
    \caption{Comparison of neural network performance on the 5nm Au experimental data with varying applied doseage in the training dataset, as sampled with three different random number generation algorithms. Results are shown for networks trained with initial learning rates of 0.01 (top) and 0.001 (bottom). In the middle column, the horizontal lines represent the median performance for selecting larger ranges of doseages in the training dataset. Error bars represent minimum and maximum performances.}
    \label{sfig:rng}
\end{figure}

For the datasets described previously, applied dosage effects were sampled as pure Poissonian noise using the MT19937 random number generator, where the random state was set by default according to the NumPy legacy generation scheme. As a test to ensure the nonlinear relationship between model performance and dose in the simulated data did not arise from quirks or correlations in the random number generation process, since large amounts of random numbers are drawn per image, we regenerated the doseage series datasets with two other high-quality psuedo-random number generators. Results, included in Figure \ref{sfig:rng}, indicate that the non-linearity is not a result of the generation scheme. These results also may imply that the variance of performance measures may be affected by the quality of random number generation when sufficiently large amounts of random numbers are sampled during data generation processes.

\section{Network Architecture}

\begin{figure}[htbp]
    \centering
    \includegraphics{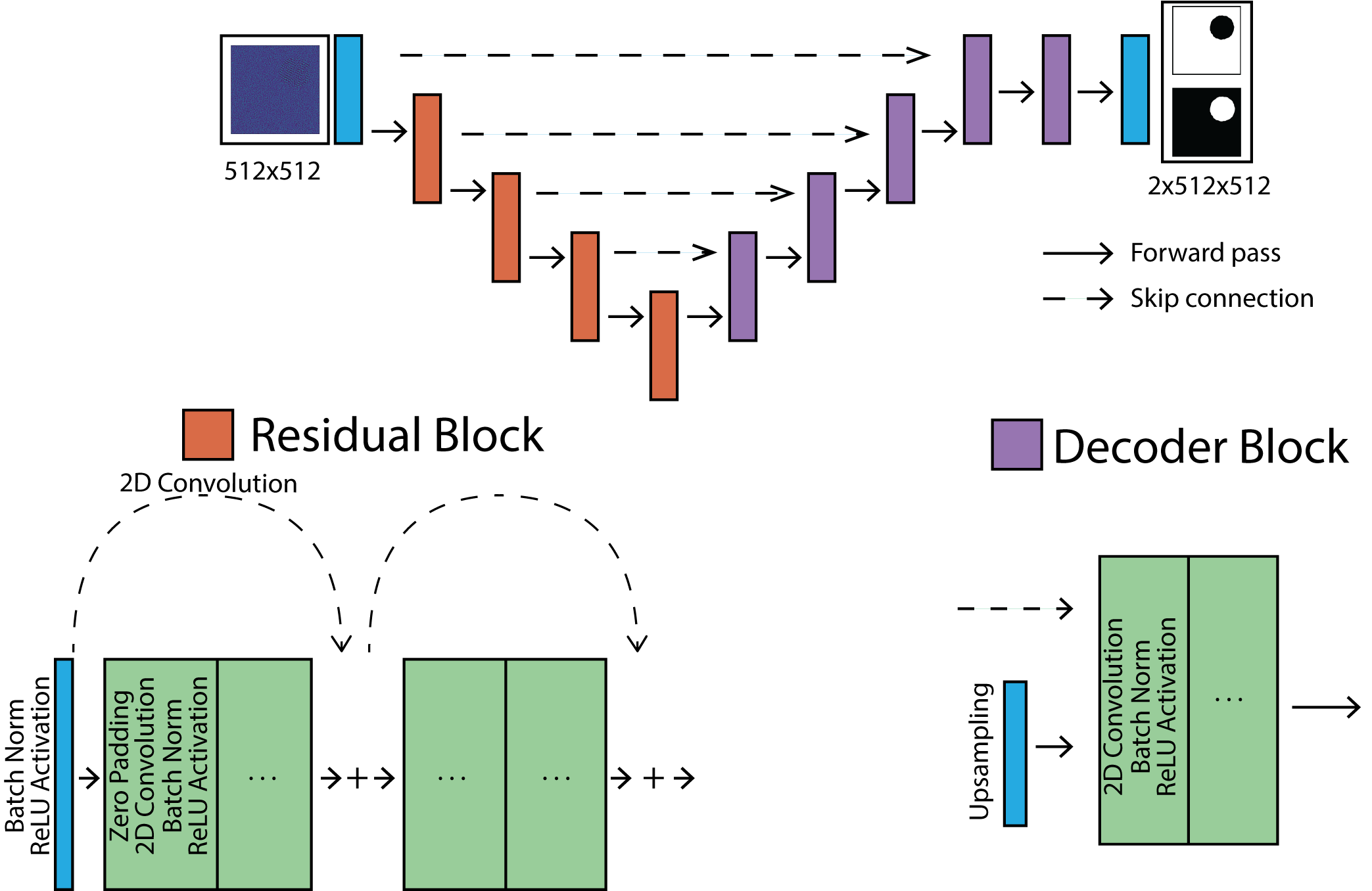}
    \caption{Architecture of the neural networks used in this study. All neural networks used a U-Net architecture \cite{ronneberger_u-net_2015} similar to the models in \cite{groschner_machine_2021}, with residual connections used in downsampling blocks and about 14M trainable parameters.}
    \label{sfig:unet_arch}
\end{figure}

\section{Additional Results}

\begin{table}[htbp]
    \centering
    \footnotesize
    \begin{tabular}{lR{1.1cm}C{2.68cm}C{2.68cm}C{2.68cm}C{2.68cm}C{2.68cm}}
    \toprule
     & & \multicolumn{5}{c}{F1-score on Exp. Data (min./median/max.)} \\
     \multicolumn{2}{c}{Training Dataset} & \multicolumn{2}{c}{TEM Operator 1} & \multicolumn{3}{c}{TEM Operator 2} \\
     \cmidrule(lr){1-2} \cmidrule(lr){3-4} \cmidrule(lr){5-7}
     Dataset & $N_{\text{Images}}$ & 5nm Au$^*$ & 2nm CdSe & 2.2nm Au & 5nm Au$^\dagger$ & 10nm Au\\
     \midrule
         Baseline &  512 &  0.422/0.634/0.723 &  0.371/0.521/0.680 &  0.478/0.665/0.770 &  0.482/0.692/0.807 &  0.543/0.740/0.836 \\
      Thermal &  512 &  0.420/0.554/0.681 &  0.304/0.521/0.625 &  0.493/0.662/0.720 &  0.491/0.703/0.771 &  0.540/0.757/0.817\\
      Noise & 512 &  0.424/0.646/0.748 &  0.339/0.473/0.650 &  0.555/0.678/0.738 &  0.617/0.733/0.811 &  0.681/0.782/0.835 \\
      Plasmonic &      512 &  0.435/0.610/0.735 &  0.434/0.586/0.642 &  0.530/0.675/0.724 &  0.569/0.750/0.809 & 
      0.636/0.790/0.835\\
      Smaller NPs &     1024 &  0.339/0.461/0.657 &  0.483/0.576/0.656 &  0.445/0.712/0.795 &  0.627/0.763/0.832 &  0.607/0.805/0.871\\
       Varying Substrate &     1024 &  0.448/0.675/0.824 &  0.473/0.568/0.637 &  0.590/0.736/0.815 &  0.658/0.771/0.851 &  0.691/0.807/0.883\\
      Optimized Au &  8000 &  0.915/0.915/0.915 &  0.648/0.648/0.648 &  0.808/0.808/0.808 &  0.909/0.909/0.909 &  0.854/0.854/0.854 \\
      Optimized mixed Au/CdSe & 8000 &  0.868/0.872/0.884 &  0.721/0.728/0.731 &  0.818/0.848/0.863 &  0.906/0.925/0.932 &  0.877/0.935/0.950 \\
      Optimized CdSe & 8000 &  0.736/0.771/0.785 &  0.722/0.737/0.747 &  0.827/0.843/0.852 &  0.850/0.863/0.873 &  0.870/0.888/0.893 \\
    \bottomrule
    \end{tabular}
    \caption{Detailed performance of summary of neural networks after training on simulated data, as measured on five experimental datasets. For each set of networks, the minimum, median, and maximum performances, as measured by F1-score, are recorded. $^*$Taken the Themis microscope. $^\dagger$Taken on the TEAM 0.5 microscope }
    \label{stab:all_perf}
\end{table}

\begin{figure}[htbp]
    \centering
\includegraphics[width=5.4in]{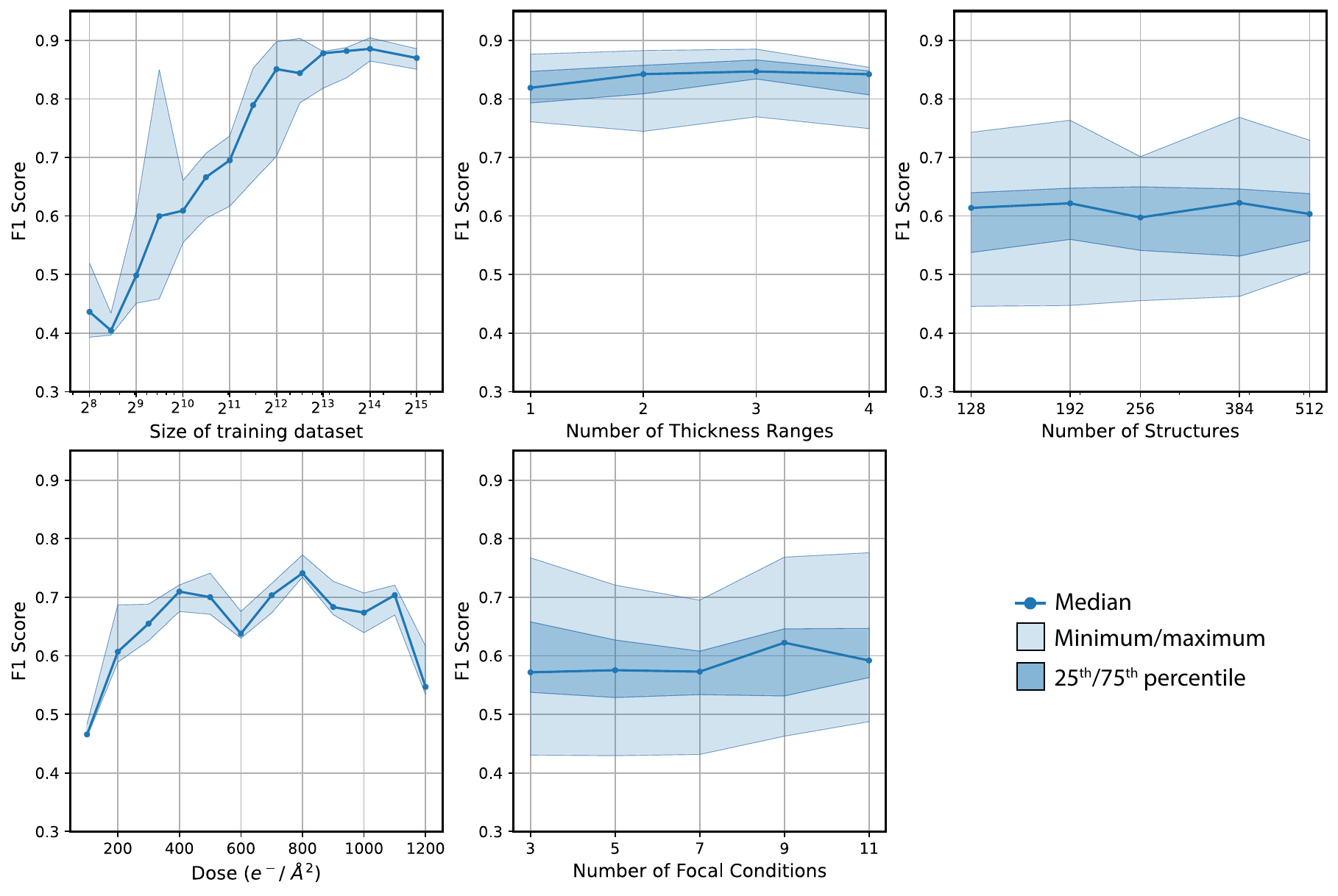}
    \caption{Summary of data effects for neural networks trained to segment nanoparticles in HRTEM micrographs. Performance is measured with respect to aspects of the dataset used to train networks, including dataset size, applied electron dose,variety of substrate thicknesses, number of focal conditions seen, and the relative proportion of focal conditions versus unique structures in the dataset. Performance is measured on micrographs of 5nm Au nanoparticles from a Themis microscope.}
    \label{sfig:large_Au}
\end{figure}
\begin{figure}[thbp]
    \centering
    \includegraphics[width=5.4in]{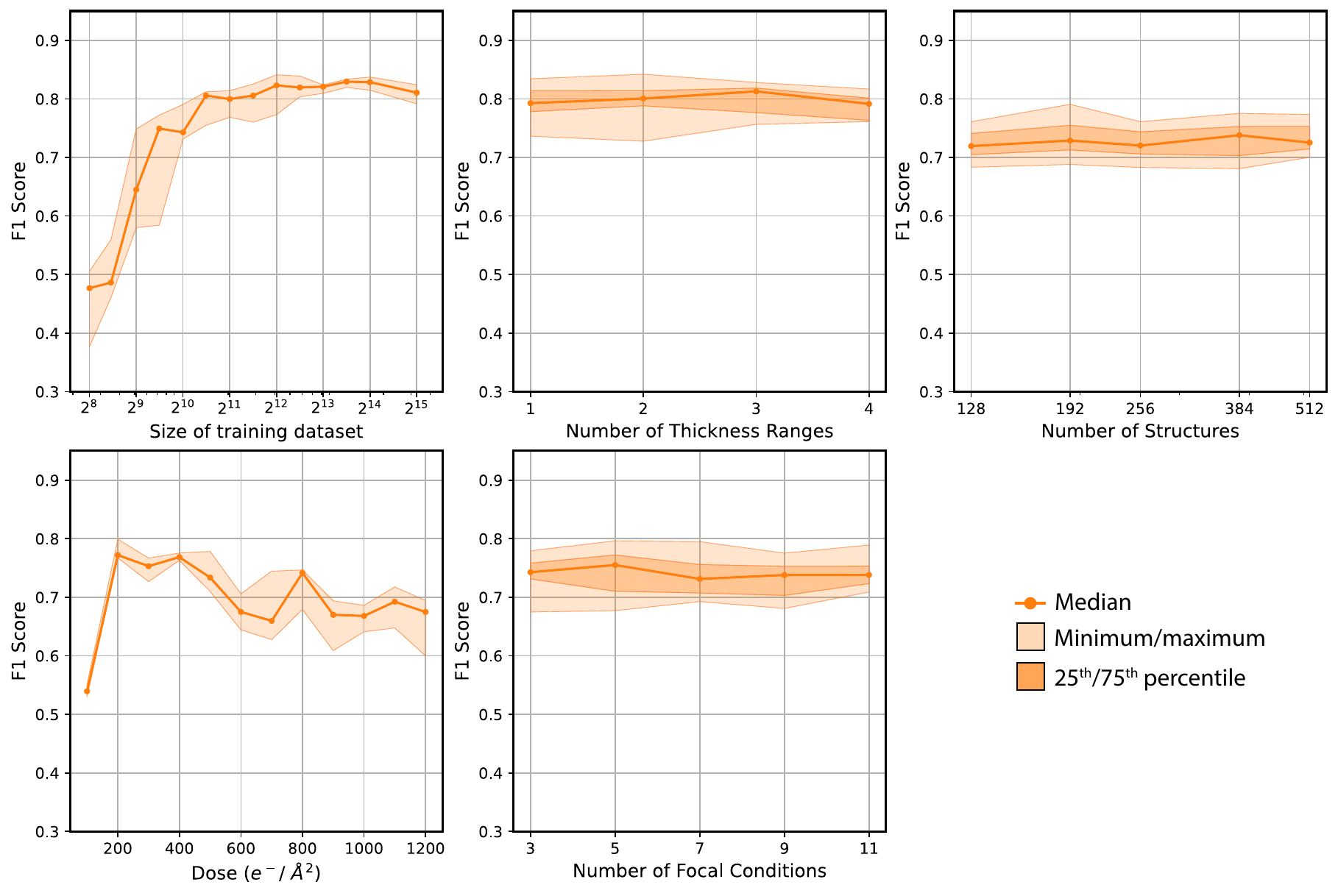}
    \caption{Summary of data effects for neural networks trained to segment nanoparticles in HRTEM micrographs. Performance is measured with respect to aspects of the dataset used to train networks, including dataset size, applied electron dose,variety of substrate thicknesses, number of focal conditions seen, and the relative proportion of focal conditions versus unique structures in the dataset. Performance is measured on micrographs of 2.2nm Au nanoparticles imaged on the TEAM 0.5 microscope at a dose rate of 423 e-/$\text{\AA}^2$.}
    \label{sfig:small_Au}
\end{figure}
\begin{figure}[bhtp]
    \centering
    \includegraphics[width=5.4in]{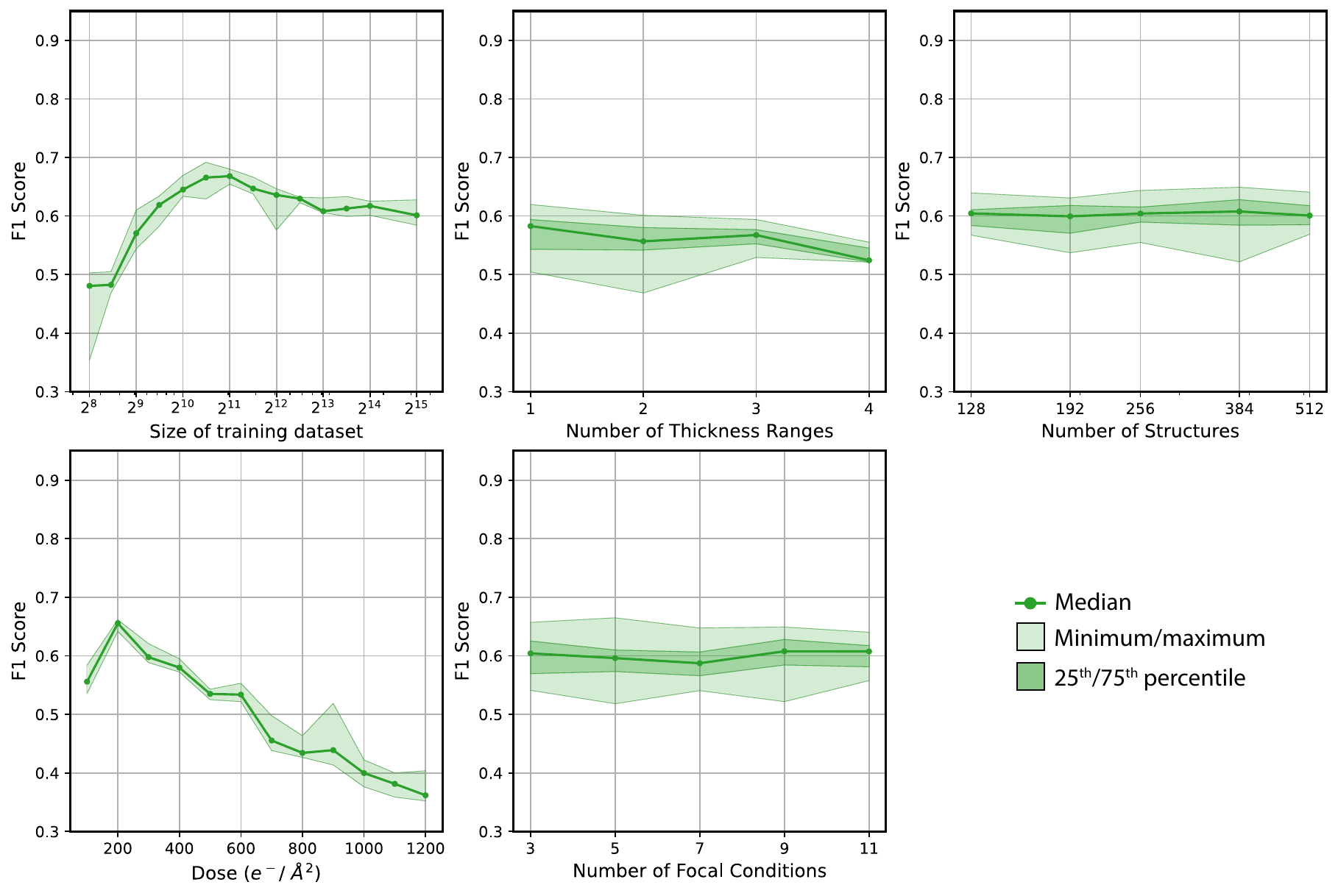}
    \caption{Summary of data effects for neural networks trained to segment nanoparticles in HRTEM micrographs. Performance is measured with respect to aspects of the dataset used to train networks, including dataset size, applied electron dose,variety of substrate thicknesses, number of focal conditions seen, and the relative proportion of focal conditions versus unique structures in the dataset. Performance is measured on micrographs of 2nm CdSe nanoparticles imaged on the TEAM 0.5 microscope at an appproximate dose rate between 200 and 600 e-/$\text{\AA}^2$.}
    \label{sfig:cdse}
\end{figure}

\begin{figure}[thbp]
    \centering
    \includegraphics[width=6in]{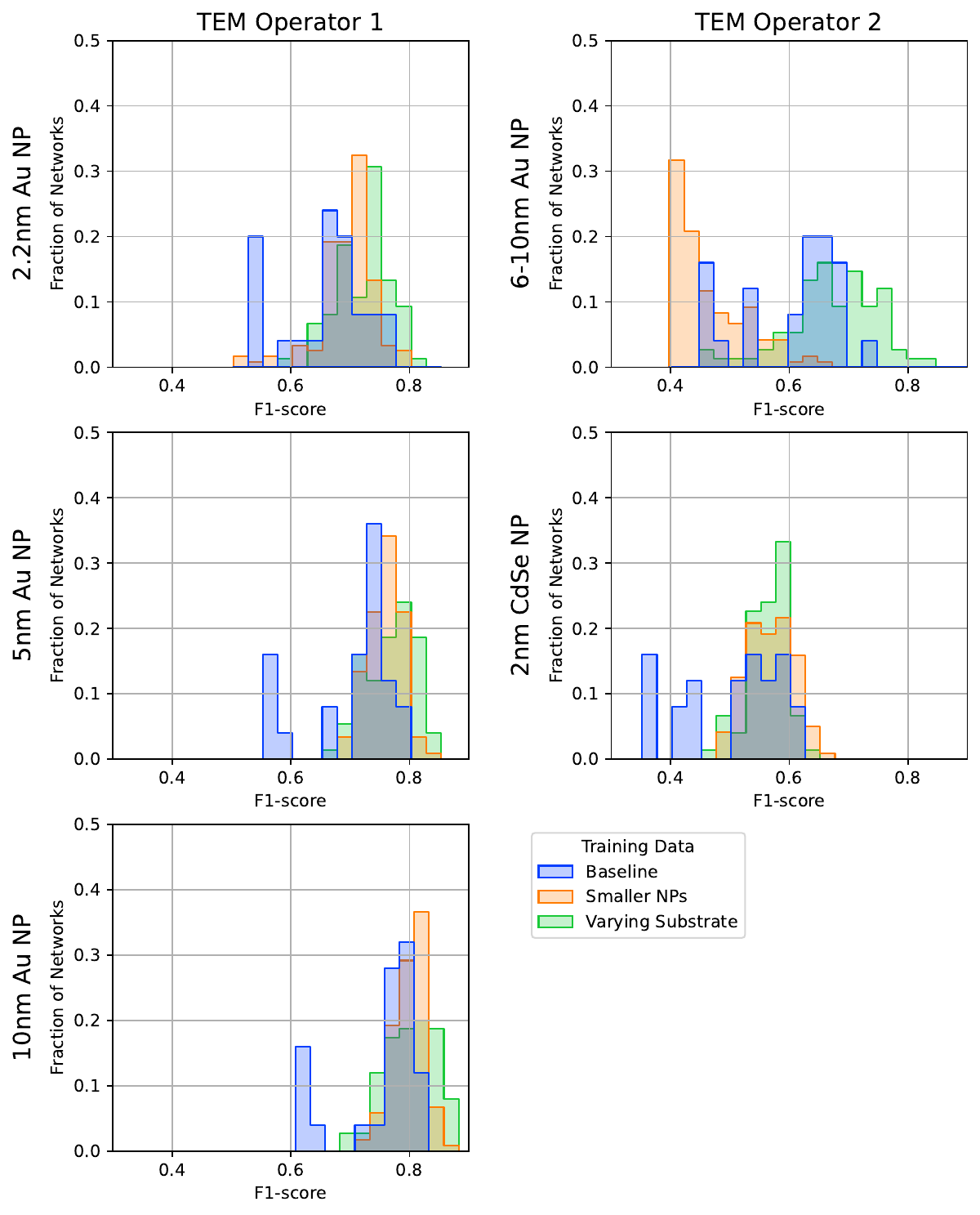}
    \caption{Comparison of three sets of neural network as trained on three different datasets and measured on five different experimental datasets from two operators.}
    \label{sfig:gen_structure}
\end{figure}

\begin{figure}[bhtp]
    \centering
    \includegraphics[width=6in]{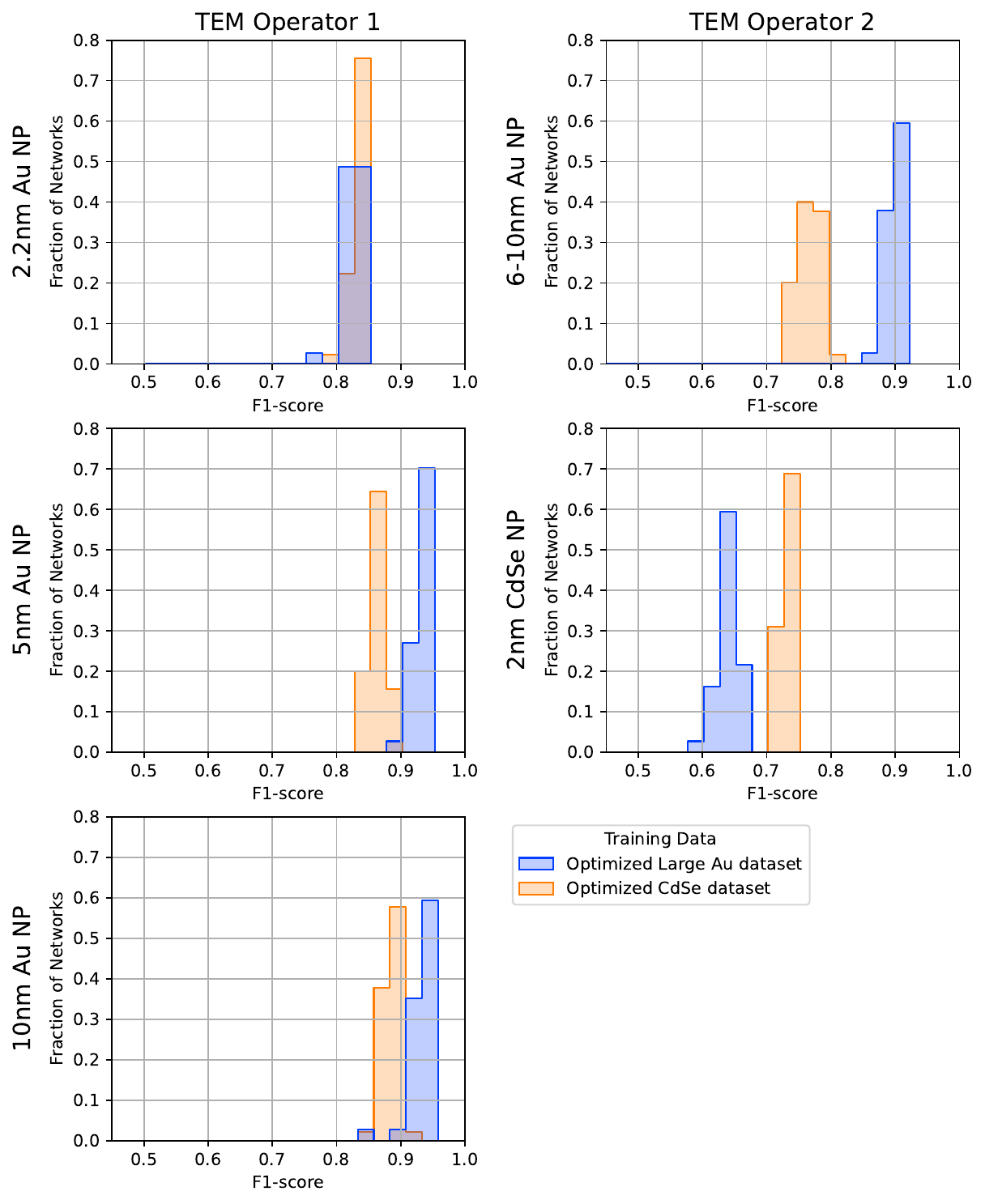}
    \caption{Comparison of two sets of neural network as trained on optimized datasets and measured on five different experimental datasets from two operators.}
    \label{sfig:gen_opt}
\end{figure}

\begin{figure}[htbp]
    \centering
    \includegraphics[width=6in]{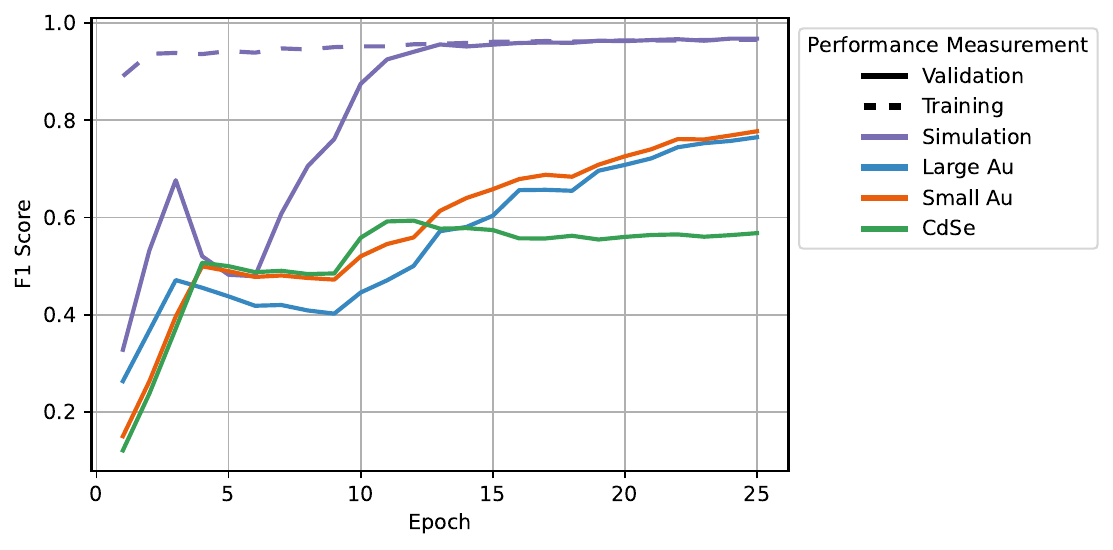}
    \caption{Characteristic training dynamics and performance over time of neural networks trained on simulated data.}
    \label{sfig:gen_dynamics}
\end{figure}

\clearpage
\section{References}
\bibliography{refs}